\documentclass{article}
\usepackage{amssymb}
\usepackage{amsfonts}
\usepackage{amsmath}
\usepackage{graphicx}

\begin{document}

\title{Bouncing ball dynamics: simple model of motion of the table and
sinusoidal motion}
\author{Andrzej Okni\'nski$^{1)}$, Bogus{\l }aw Radziszewski$^{2)}$ \\
Kielce University of Technology, 25-314 Kielce, Poland$^{1)}$\\
Collegium Mazovia Innovative University, 08-110 Siedlce, Poland$^{2)}$\\
}
\maketitle

\begin{abstract}
Nonlinear dynamics of a bouncing ball moving vertically in a gravitational
field and colliding with a moving limiter is considered and the Poincar\'{e}
map, describing evolution from an impact to the next impact, is described.
Displacement of the table is approximated in one period by four cubic
polynomials. Results obtained for this model are used to elucidate dynamics
of the standard model of bouncing ball with sinusoidal motion of the limiter.
\end{abstract}

\section{Introduction}

In the present paper we study dynamics of a small ball moving vertically in
a gravitational field and impacting with a periodically moving limiter (a
table). This model belongs to the field of nonsmooth and nonlinear dynamical
systems \cite{diBernardo2008,Luo2006,Awrejcewicz2003,Filippov1988}. In such
systems nonstandard bifurcations such as border-collisions and grazing
impacts leading often to complex chaotic motions are typically present. It
is important that nonsmooth systems have many applications in technology 
\cite{Stronge2000,Mehta1994,Knudsen1992,Wiercigroch2008,Awrejcewicz2007}.

Impacting systems studied in the literature can be divided into three main
classes: bouncing ball models \cite{Holmes1982,Luo1996,Luo2009a}, impacting
oscillators \cite{Nordmark2001} and impacting pendulums \cite%
{Lenci2000,Awrejcewicz2007}, see also \cite{diBernardo2008}. In dynamics
with impacts it is usually difficult or even impossible to solve nonlinear
equation for an instant of the next impact. For example, in the bouncing
ball models the table's motion has been usually assumed in sinusoidal form,
cf. \cite{Luo2009a} and references therein. This choice of the limiter's
motion leads indeed to nontractable nonlinear equation for time of the next
impact. To tackle this problem we proposed a sequence of models in which
periodic motion of the table is assumed (in one period of limiter's motion)
as a low-order polynomial of time \cite{AOBR2010}. It is thus possible to
approximate the sinusoidal motion of the table more and more exactly and
conduct analytical computations. Carrying out this plan we have studied
several such models with linear, quadratic and cubic polynomials \cite%
{AOBR2009,AOBR2011,AOBR2012a,AOBR2012b}.

In the present work we conduct analytical and numerical investigations of
the model in which sinusoidal displacement of the table is approximated in
one period by four cubic polynomials. We shall refer to this model as $%
\mathcal{M}_{C}$. Simultaneously, we study the standard dynamics of bouncing
ball with sinusoidal motion of the limiter, referred to as $\mathcal{M}_{S}$%
. We hope that rigorous results obtained for the model $\mathcal{M}_{C}$\
cast light on dynamics of $\mathcal{M}_{S}$. It should be stressed that
results obtained for the model $\mathcal{M}_{S}$ can be compared with
experimental studies, see \cite{Pieranski1983,Tufillaro1986,Celaschi1987}
for the early papers, summarized in \cite{Pieranski1994}, and \cite%
{Eichwald2010} for recent work.

The paper is organized as follows. In Section 2 a one dimensional dynamics
of a ball moving in a gravitational field and colliding with a table is
reviewed and the corresponding Poincar\'{e} map is constructed and models of
the limiter's motion $\mathcal{M}_{C}$ and $\mathcal{M}_{S}$ are defined.
Bifurcation diagrams are computed for $\mathcal{M}_{C}$ and $\mathcal{M}_{S}$%
. In Sections 3, 4 and 5 a combination of analytical and numerical approach
is used to investigate selected problems of dynamics in models $\mathcal{M}%
_{C}$ and $\mathcal{M}_{S}$. More exactly, fixed points and their stability
are discussed in Section 3, birth of low velocity $n$-cycles is investigated
in Section 4 and birth of high velocity $3$-cycles is studied in Section 5.
In Section 6 the case of $N$ impacts in one interval of the limiter's motion
is studied for the model $\mathcal{M}_{C}$. We summarize our results in the
last Section.

\section{Bouncing ball: a simple motion of the table}

Let a ball moves vertically in a constant gravitational field and collides
with a periodically moving table. We treat the ball as a material point and
assume that the limiter's mass is so large that its motion is not affected
at impacts. Dynamics of the ball from an impact to the next impact can be
described by the following Poincar\'{e} map in nondimensional form \cite%
{AOBR2007} (see also Ref. \cite{Luo1996} where analogous map was derived
earlier and Ref. \cite{Luo2009a} for generalizations of the bouncing ball
model): 
\begin{subequations}
\label{TV}
\begin{eqnarray}
\gamma Y\left( T_{i+1}\right) &=&\gamma Y\left( T_{i}\right) -\Delta
_{i+1}^{2}+\Delta _{i+1}V_{i},  \label{T} \\
V_{i+1} &=&-RV_{i}+2R\Delta _{i+1}+\gamma \left( 1+R\right) \dot{Y}\left(
T_{i+1}\right) ,  \label{V}
\end{eqnarray}%
where $T_{i}$ denotes time of the $i$-th impact and $V_{i}$ is the
corresponding post-impact velocity while $\Delta _{i+1}\equiv T_{i+1}-T_{i}$%
. The parameters $\gamma $, $R$ are a nondimensional acceleration and the
coefficient of restitution, $0\leq R<1$ \cite{Stronge2000}, respectively and
the function $Y\left( T\right) $ represents the limiter's motion. The
limiter's motion has been typically assumed in sinusoidal form, $%
Y_{S}(T)=\sin (2\pi T)$. Equations (\ref{TV}) and $Y=Y_{S}$ lead to the
model $\mathcal{M}_{S}$. This choice of limiter's motion leads to serious
difficulties in solving the first of Eqns.(\ref{TV}) for $T_{i+1}$, thus
making analytical investigations of dynamics hardly possible. Accordingly,
we have decided to simplify the limiter's periodic motion to make (\ref{T})
solvable. The function $Y_{C}\left( T\right) $: 
\end{subequations}
\begin{equation}
Y_{C}\left( T\right) =\left\{ 
\begin{array}{cc}
f_{1}\left( T\right) , & 0\leq \hat{T}<\frac{1}{4} \\ 
f_{2}\left( T\right) , & \frac{1}{4}\leq \hat{T}<\frac{1}{2} \\ 
f_{3}\left( T\right) , & \frac{1}{2}\leq \hat{T}<\frac{3}{4} \\ 
f_{4}\left( T\right) , & \frac{3}{4}\leq \hat{T}\leq 1%
\end{array}%
\right.  \label{Yc2}
\end{equation}

\begin{subequations}
\label{C2B}
\begin{eqnarray}
\hspace{-21pt}f_{1}\left( T\right) \! &=&\!\left( 32\pi -128\right) \hat{T}%
^{3}+\left( -16\pi +48\right) \hat{T}^{2}+2\pi \hat{T}  \label{C2b1} \\
\hspace{-21pt}f_{2}\left( T\right) \! &=&\!\left( 128-32\pi \right) \hat{T}%
^{3}+\left( -144+32\pi \right) \hat{T}^{2}+\left( 48-10\pi \right) \hat{T}%
-4+\pi  \label{C2b2} \\
\hspace{-21pt}f_{3}\left( T\right) \! &=&\!\left( 128-32\pi \right) \hat{T}%
^{3}+\left( -240+64\pi \right) \hat{T}^{2}+\left( 144-42\pi \right) \hat{T}%
-28+9\pi  \label{C2b3} \\
\hspace{-21pt}f_{4}\left( T\right) \! &=&\!\left( 32\pi -128\right) \hat{T}%
^{3}+\left( 336-80\pi \right) \hat{T}^{2}+\left( -288+66\pi \right) \hat{T}%
+80-18\pi  \label{C2b4}
\end{eqnarray}%
approximates $Y_{S}=\sin (2\pi T)$ on the intervals $\left[ k,\,k+1\right] $%
, $k=0,1,\ldots ,$ with $\hat{T}=T-\left\lfloor T\right\rfloor $, where $%
\left\lfloor x\right\rfloor $ is the floor function -- the largest integer
less than or equal to $x$. The model $\mathcal{M}_{C}$ consists of equations
(\ref{TV}), (\ref{Yc2}), (\ref{C2B}) with control parameters $R$, $\gamma $. We shall
also need velocities of the limiter, defined as $g_{i}\left( T\right) 
\overset{df}{=}\frac{d}{dt}f_{i}\left( T\right) $, $i=1,\ldots ,4$.

\begin{figure}[th!]
\center 
\includegraphics[width=10cm, height=8cm]{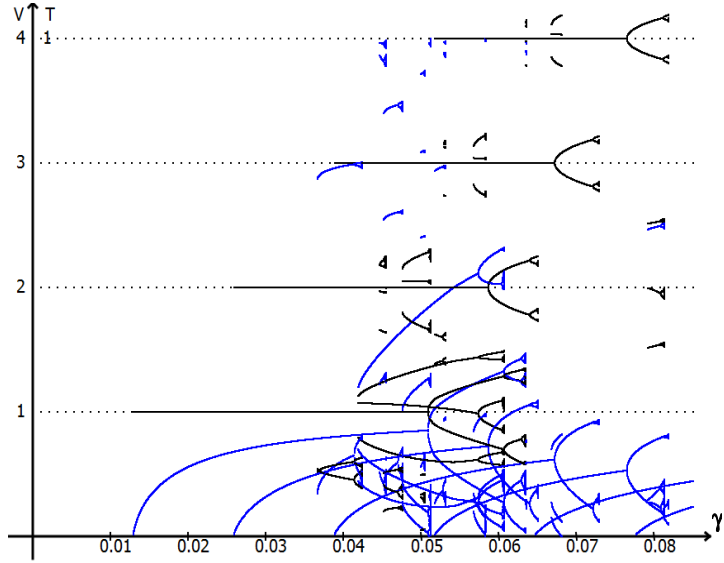}
\caption{Bifurcation diagram for the model $\mathcal{M}_{C}$, $R=0.85$.}
\end{figure}

In Fig. 1 above we show the bifurcation diagram with impact times (blue) and
velocities (black) versus $\gamma $ computed for growing $\gamma $ and $%
R=0.85$. It follows that dynamical system $\mathcal{M}_{C}$ has
several attractors: two fixed points which after one period doubling give
rise to chaotic bands and two other fixed points which go to chaos via
period doubling scenario. There are also several small attractors. We shall
investigate some of these attractors in the next Section combining
analytical and numerical approach (general analytical conditions for birth
of new modes of motion were given in \cite{Luo2009b}).

We show below the corresponding bifurcation diagram for the sinusoidal
motion.

\begin{figure}[th!]
\center 
\includegraphics[width=10cm, height=8cm]{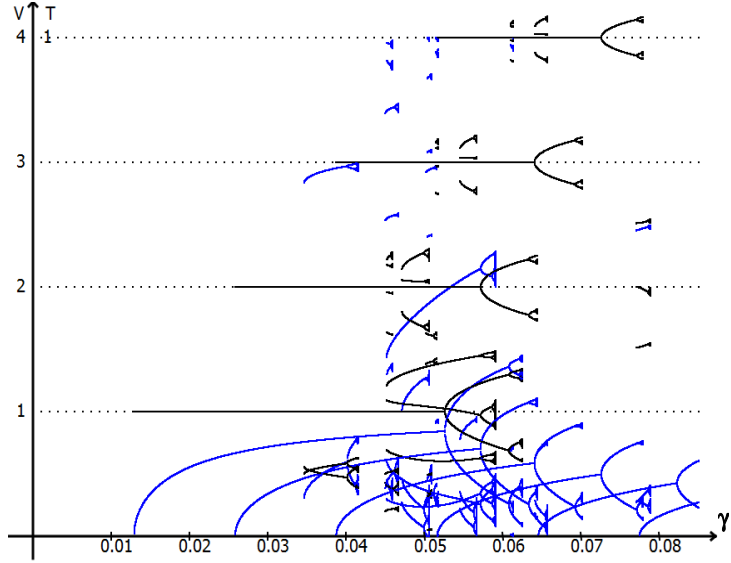}
\caption{Bifurcation diagram for the model $\mathcal{M}_{S}$, $R=0.85$.}
\end{figure}

Similarity of Figs. 1, 2 suggests that analytical results obtained for model 
$\mathcal{M}_{C}$ shed light on the problem of sinusoidal motion, $\mathcal{M%
}_{S}$.

\section{Fixed points and their stability}

We shall first study periodic solutions of the model $\mathcal{M}_{C}$ with
one impact per $k$ periods and $T\in \left( 0,\ \frac{1}{4}\right) $ since%
{\Huge \ }it is suggested by the bifurcation diagram that they are stable.
Such states have to fulfill the following conditions: 
\end{subequations}
\begin{equation}
V_{n+1}=V_{n}\equiv V_{\ast }^{\left( k/1\right) },\ T_{n+1}=T_{n}+k\equiv
T_{\ast }^{\left( k/1\right) }+k\qquad \left( k=1,2,\ldots \right) ,
\label{cond1a}
\end{equation}%
where:%
\begin{equation}
T_{\ast }^{\left( k/1\right) }\in \left( 0,\ \tfrac{1}{4}\right) ,\ V_{\ast
}^{\left( k/1\right) }>\gamma \dot{Y}_{c_{1}}\left( T_{\ast }^{\left(
k/1\right) }\right) .  \label{cond1b}
\end{equation}

The demanded (stable) solution is given by 
\begin{subequations}
\label{SOL1}
\begin{eqnarray}
T_{\ast \left( s\right) }^{\left( k/1\right) } &=&\tfrac{\pi -3}{6\left( \pi
-4\right) }-\tfrac{1}{24\left( \pi -4\right) }\sqrt{4\left( \pi -6\right)
^{2}+6B\left( \pi -4\right) },\quad \left( B=\tfrac{k}{\gamma }\tfrac{1-R}{%
1+R}\right)  \label{sol1a} \\
V_{\ast }^{\left( k/1\right) } &=&k.  \label{sol1b}
\end{eqnarray}%
Since $T_{\ast }\in \left[ 0,\ 1\right] $ we demand that $T_{\ast }>0$ and
it follows from (\ref{sol1a}) that physical solution appears for lower
critical value $\gamma >\gamma _{cr_{1}}^{\left( k/1\right) }$ where: 
\end{subequations}
\begin{equation}
\gamma _{cr_{1},C}^{\left( k/1\right) }=\tfrac{k}{2\pi }\tfrac{1-R}{1+R}.
\label{critical1a}
\end{equation}

We have checked by stability analysis that the solution (\ref{sol1a}), (\ref%
{sol1b}) is stable for $\gamma >\gamma _{cr_{1},C}^{\left( k/1\right) }$,
i.e. when it is physically acceptable. To determine upper critical value of $%
\gamma $ when dynamics looses stability we put into (\ref{TV}):%
\begin{eqnarray}
T_{i} &=&T_{\ast \left( s\right) }^{\left( k/1\right) }+\varepsilon _{i},\
T_{i+1}=T_{\ast \left( s\right) }^{\left( k/1\right) }+k+\varepsilon _{i+1},
\label{stab1a} \\
V_{i} &=&V_{\ast }+\mu _{i}=k+\mu _{i},\ V_{i+1}=V_{\ast }+\mu _{i+1}=k+\mu
_{i+1},  \label{stab1b}
\end{eqnarray}%
with $Y\left( T\right) $ given by (\ref{C2B}), and keep only terms linear in
perturbations $\varepsilon _{i}$, $\varepsilon _{i+1}$, $\mu _{i}$, $\mu
_{i+1}$ of the fixed point to get:

\begin{equation}
\left( 
\begin{array}{c}
\varepsilon _{i+1} \\ 
\mu _{i+1}%
\end{array}%
\right) =\left( 
\begin{array}{ll}
1 & \ \tfrac{k}{\gamma f_{1}\left( T_{\ast }\right) +k} \\ 
\gamma \left( 1+R\right) g_{1}\left( T_{\ast }\right) & \ k\tfrac{2R+\gamma
\left( 1+R\right) g_{1}\left( T_{\ast }\right) }{\gamma f_{1}\left( T_{\ast
}\right) +k}-R%
\end{array}%
\right) \left( 
\begin{array}{c}
\varepsilon _{i} \\ 
\mu _{i}%
\end{array}%
\right)  \label{lin1}
\end{equation}%
where $T_{\ast }\equiv T_{\ast \left( s\right) }^{\left( k/1\right) }$, $%
f_{1}\left( T\right) $ is given by (\ref{C2b1}) and $g_{1}\left( T\right) =%
\frac{d}{dT}f_{1}\left( T\right) $.

Since the characteristic polynomial is:%
\begin{equation}
\begin{array}{l}
X^{2}+\alpha X+\beta =0 \\ 
\alpha =4\sqrt{4\left( \pi -6\right) ^{2}+6k\left( \pi -4\right) \frac{1-R}{%
\gamma \left( 1+R\right) }}\left( 1+R\right) ^{2}\gamma -R^{2}-1 \\ 
\beta =R^{2}%
\end{array}
\label{char1}
\end{equation}%
application of the Shur-Cohn criterion (\cite{Jury1974}):%
\begin{eqnarray}
\beta &<&1  \label{SC1} \\
\left\vert \alpha \right\vert &<&\beta +1  \notag
\end{eqnarray}%
leads finally to the localization of the fixed points (\ref{SOL1}), $\gamma
_{cr_{1},C}^{\left( k/1\right) }<\gamma <\gamma _{cr_{2},C}^{\left(
k/1\right) }$, with:%
\begin{equation}
\gamma _{cr_{2},C}^{\left( k/1\right) }=\tfrac{6k\left( \pi -4\right) \left(
R^{2}-1\right) +\sqrt{36k^{2}\left( \pi -4\right) ^{2}\left( 1-R^{2}\right)
^{2}+4\left( \pi -6\right) ^{2}\left( 1+R^{2}\right) ^{2}}}{8\left( \pi
-6\right) ^{2}\left( 1+R\right) ^{2}},\quad R<1.  \label{critical1b}
\end{equation}

In Fig. 3 stability regions in $\left( R,\gamma \right) $ plane for the $%
\mathcal{M}_{C}$ model are shown. In the case of the model $\mathcal{M}_{S}$
we have:

\begin{eqnarray}
\gamma _{cr_{2},S}^{\left( k/1\right) } &=&\gamma _{cr_{2},C}^{\left(
k/1\right) },  \label{critical1a_sine} \\
\gamma _{cr_{2},S}^{\left( k/1\right) } &=&\tfrac{\sqrt{k^{2}\pi
^{2}(1-R^{2})^{2}+4(1+R^{2})^{2}}}{2\pi ^{2}(1+R)^{2}},\quad R<1,
\label{critical1b_sine}
\end{eqnarray}%
see \cite{AOBR2007} $($note that in \cite{AOBR2007} we used $Y(T)=\sin (T)$
rather than $Y_{S}(T)=\sin (2\pi T)$ and it follows that all values of the
control parameter $\lambda $ must be rescaled, $\gamma =\frac{\lambda }{%
\left( 2\pi \right) ^{2}}$) and stability regions are very similar to those
of model $\mathcal{M}_{C}$, cf. Fig. 4.

\begin{figure}[th!]
\center
\includegraphics[width=10cm,height=8cm]{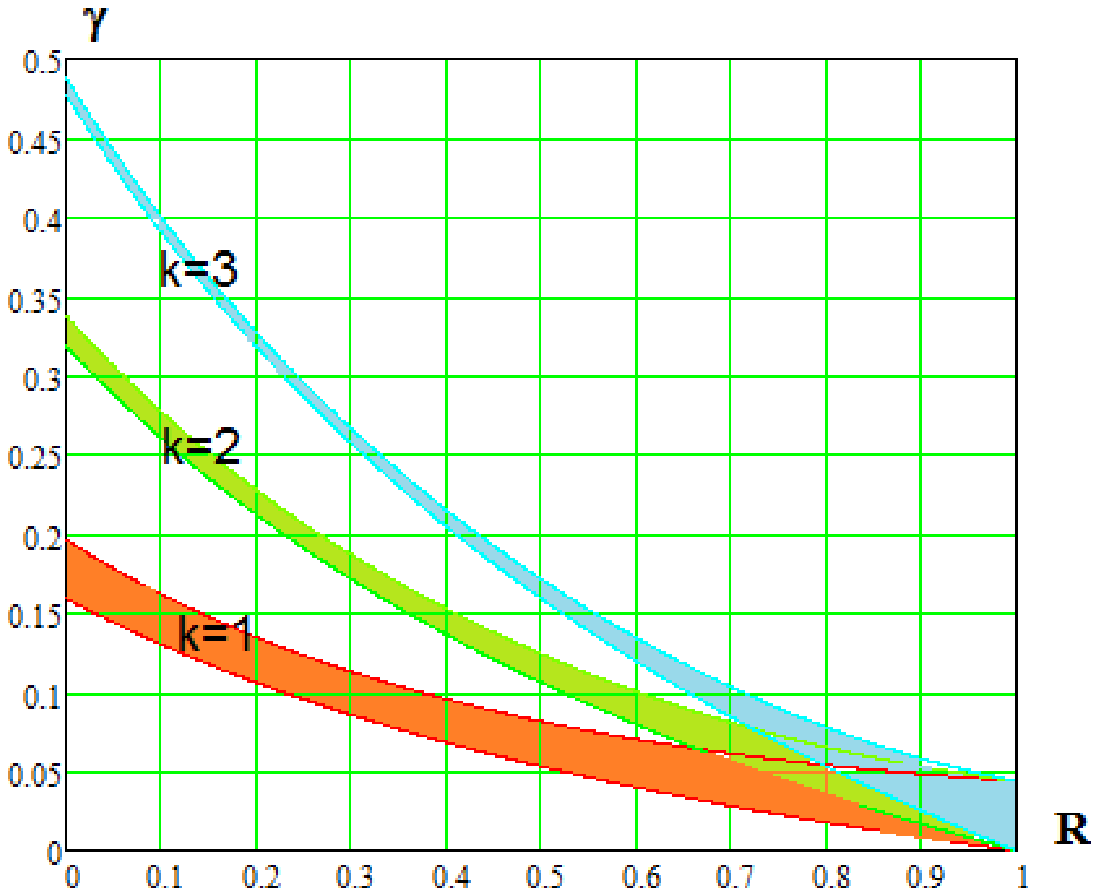}
\caption{Stability regions in the $\left( R,\protect\gamma \right) $ plane, model $\mathcal{M}_{C}$.} 
\center
\includegraphics[width=10cm, height=8cm]{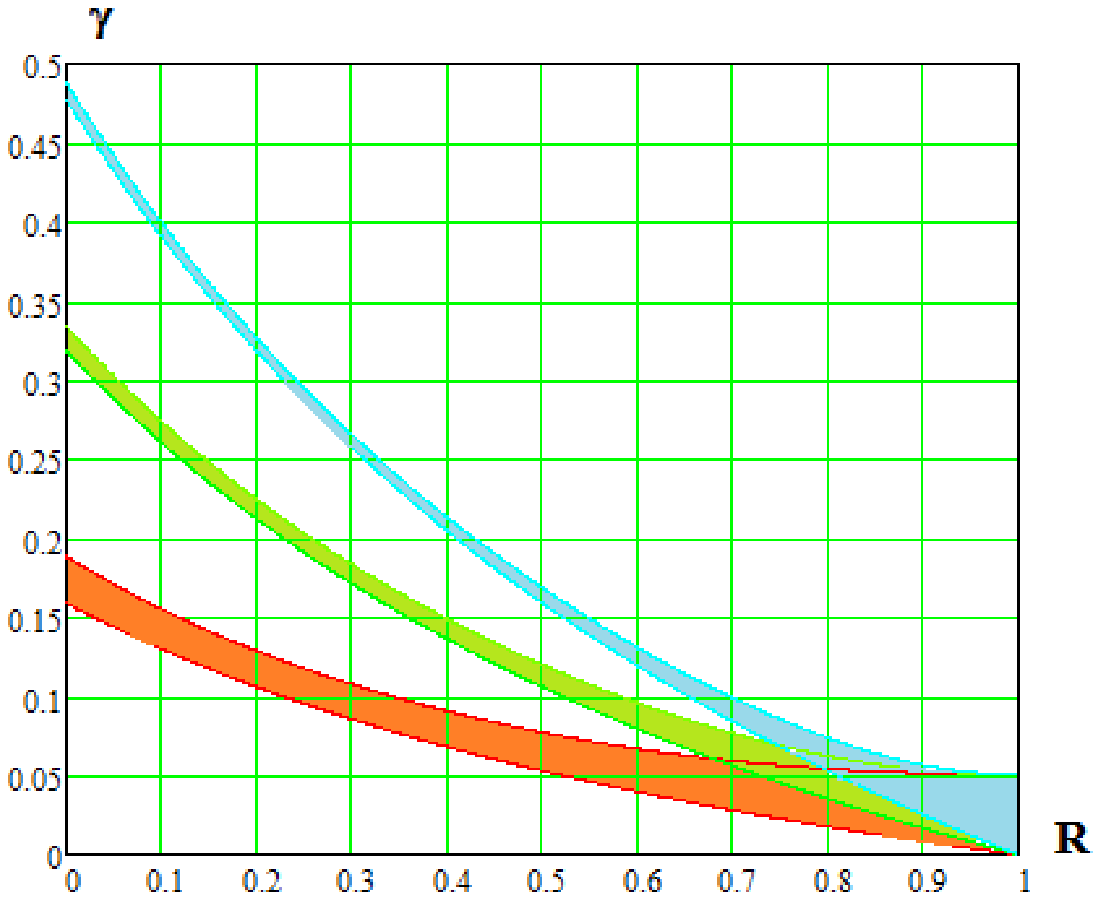}
\caption{Stability regions in the $\left( R,\protect\gamma \right) $ plane, model $\mathcal{M}_{S}$.}
\label{F34}
\end{figure}

\section{Birth of low velocity $k$ - cycles}

In this Subsection we shall study birth of low velocity $k$ - cycles which
can be seen in the bifurcation diagrams, Figs. 1, 2, for $\gamma >0.03$ and $%
V<1$. In the case of such cycles $T_{1},T_{2},\ldots ,T_{k}\in \left(
0,1\right) $ and $T_{k+1}-1=T_{1}$. Of course, it is possible to follow
periodic orbits backwards, i.e. iterating the map (\ref{TV}) until the
convergence to the $k$ - cycle is achieved for some initial condition and
some $\gamma $. Then the value of $\gamma $ is decreased (slightly) and the
map is iterated again (until convergence is obtained) with the previously
computed $k$ - cycle as the initial condition. This method, although leads
to determination of the critical value of $\gamma $ at which the $k$ - cycle
disappears for decreasing $\gamma $ (or is born for growing $\gamma $) but
is time-consuming \ and not very effective due to very poor convergence near
the threshold.

On the other hand, analytical conditions for birth of $k$ - cycles are found
below. In what follows theorems about differentiation of implicit functions 
\cite{Krantz2003} will turn out useful since Eqn. (\ref{T}) defines $T_{i+1}$
implicitly. Consider equation:%
\begin{equation}
F\left( T_{1},T_{2}\right) =0,  \label{Impl1a}
\end{equation}%
which defines dependence of, say, $T_{2}$ on $T_{1}$, see \cite{Krantz2003}
where necessary and sufficient assumptions are given. Then it follows from
implicit function theorem that:%
\begin{equation}
\frac{dT_{2}}{dT_{1}}=-\frac{F_{1}^{\prime }}{F_{2}^{\prime }}.
\label{Impl1b}
\end{equation}%
where $F_{1}^{\prime }\equiv \frac{\partial F}{\partial T_{1}}$, $%
F_{2}^{\prime }\equiv \frac{\partial F}{\partial T_{2}}$.

In a more complicated case, equations:

\begin{equation}
F\left( T_{1},T_{2},T_{3}\right) =0,\quad G\left( T_{1},T_{2},T_{3}\right)
=0,  \label{Impl2a}
\end{equation}%
define $T_{2}$ and $T_{3}$ as functions of $T_{1}$ under appropriate
assumptions. We can now compute derivatives with respect to $T_{1}$ as \cite%
{Krantz2003}:%
\begin{equation}
\tfrac{\partial T_{2}}{\partial T_{1}}=-\tfrac{\det \left( 
\begin{array}{cc}
F_{1}^{\prime } & G_{1}^{\prime } \\ 
F_{3}^{\prime } & G_{3}^{\prime }%
\end{array}%
\right) }{\det \left( 
\begin{array}{cc}
F_{2}^{\prime } & G_{2}^{\prime } \\ 
F_{3}^{\prime } & G_{3}^{\prime }%
\end{array}%
\right) },\quad \tfrac{\partial T_{3}}{\partial T_{1}}=-\tfrac{\det \left( 
\begin{array}{cc}
F_{2}^{\prime } & G_{2}^{\prime } \\ 
F_{1}^{\prime } & G_{1}^{\prime }%
\end{array}%
\right) }{\det \left( 
\begin{array}{cc}
F_{2}^{\prime } & G_{2}^{\prime } \\ 
F_{3}^{\prime } & G_{3}^{\prime }%
\end{array}%
\right) },  \label{Impl2b}
\end{equation}%
with $F_{1}^{\prime }\equiv \frac{\partial F}{\partial T_{1}}$, $%
F_{2}^{\prime }\equiv \frac{\partial F}{\partial T_{2}}$, $F_{3}^{\prime
}\equiv \frac{\partial F}{\partial T_{3}}$ and analogous notation for $%
G_{i}^{\prime }$, $i=1,2,3$.

\subsection{Low velocity $2$ - cycle in the model $\mathcal{M}_{C}$}

Numerical tests show that a $2$ - cycle fulfilling conditions $T_{1}\in
\left( 0,\ \frac{1}{4}\right) ,\ T_{2}\in \left( \frac{1}{2},\ \frac{3}{4}%
\right) $ and $T_{3}=T_{1}+1$\ is stable. This $2$ - cycle can be seen in
the bifurcation diagram in Fig. 1 for $\gamma \gtrsim 0.0366$ and $%
V_{1}\cong 0.51$, $V_{2}\cong 0.55$ ($R=0.85$). Equations to determine $%
T_{1},\ T_{2}$ and $V_{1},\ V_{2}$ are shown below:%
\begin{equation}
\begin{array}{rll}
\gamma f_{3}\left( T_{2}\right) & = & \gamma f_{1}\left( T_{1}\right)
-\left( T_{2}-T_{1}\right) ^{2}+\left( T_{2}-T_{1}\right) V_{1} \\ 
V_{2} & = & -RV_{1}+2R\left( T_{2}-T_{1}\right) +\gamma \left( 1+R\right)
g_{3}\left( T_{2}\right) \\ 
\gamma f_{1}\left( T_{3}-1\right) & = & \gamma f_{3}\left( T_{2}\right)
-\left( T_{3}-T_{2}\right) ^{2}+\left( T_{3}-T_{2}\right) V_{2} \\ 
V_{3} & = & -RV_{2}+2R\left( T_{3}-T_{2}\right) +\gamma \left( 1+R\right)
g_{1}\left( T_{3}-1\right) \\ 
T_{3} & = & T_{1}+1 \\ 
V_{3} & = & V_{1}%
\end{array}
\label{2-cycle-1}
\end{equation}%
where $f_{i}\left( T\right) $'s and $g_{i}\left( T\right) $'s\ are defined
in Eqn. (\ref{C2B}) and the text below.

We were able to simplify Eqns. (\ref{2-cycle-1}) significantly obtaining
equation for $\Delta \equiv T_{2}-T_{1}$ only:%
\begin{equation}
F\left( \Delta \right) =\sum\nolimits_{j=0}^{9}d_{j}\Delta ^{j}=0,
\label{Delta}
\end{equation}%
\-where $d_{i}$'s are given in the Appendix. Numerical computations suggest
that the $2$ - cycle appears for $\gamma =\gamma _{cr,C}^{\left( 2\right) }$
and fixed $R$, where $\gamma _{cr,C}^{\left( 2\right) }$ is a critical
value, as a double (and stable) solution of Eqns. (\ref{2-cycle-1}). For $%
\gamma >\gamma _{cr,C}^{\left( 2\right) }$ there are two real solutions, one
stable (seen in the bifurcation diagram) and another unstable. On the other
hand, for $\gamma <\gamma _{cr,C}^{\left( 2\right) }$ the solutions are
complex conjugated and thus unphysical. Moreover, at $\gamma =\gamma
_{cr,C}^{\left( 2\right) }$ the stability matrix has unit eigenvalue.
Therefore this is a tangent (saddle-node) bifurcation, see \cite{Peitgen1992}
for elementary discussion of the tangent bifurcation in the logistic map
when the $3$ - cycle is born. All other cycles discussed in our paper are
also born in tangent bifurcation.

To determine critical value of the parameter $\gamma $ let us note that
double solution of the polynomial equation (\ref{Delta}) is also the
solution of $G\left( \Delta \right) =0$ where $G\left( \Delta \right) =\frac{%
d}{d\Delta }F\left( \Delta \right) $. For example, solving for $R=0.85$ the
system of equations:%
\begin{eqnarray}
F\left( \Delta \right) &=&\sum\nolimits_{j=0}^{9}d_{j}\Delta ^{j}=0,
\label{cr1} \\
G\left( \Delta \right) &=&\sum\nolimits_{j=1}^{j}jd_{j}\Delta ^{j-1}=0,
\label{cr2}
\end{eqnarray}%
we get $\gamma _{cr,C}^{\left( 2\right) }=0.036\,617\,052\,682\,892\,250\,62$%
, $\Delta _{cr}=0.634\,279\,960\,677\,747\,355\,95$ (and many other,
unphysical solutions) in perfect agreement with numerical computations, see
also Fig. 1.

Alternatively, we can use implicit function theorem. Solving the second and
fourth equations in (\ref{2-cycle-1}) for $V_{1}$, $V_{2}$ we get%
\begin{equation}
\begin{array}{l}
V_{1}=\dfrac{\gamma \left( 1+R\right) \left( Rg_{3}\left( T_{2}\right)
-g_{1}\left( T_{1}\right) \right) +2R\left( 1+R\right) \left(
T_{2}-T_{1}\right) -2R}{-1+R^{2}} \\ 
V_{2}=\dfrac{\gamma \left( 1+R\right) \left( Rg_{1}\left( T_{1}\right)
-g_{3}\left( T_{2}\right) \right) -2R\left( 1+R\right) \left(
T_{2}-T_{1}\right) +2R^{2}}{-1+R^{2}}%
\end{array}
\label{V1V2}
\end{equation}%
and%
\begin{equation}
\hspace{-6pt}%
\begin{array}{l}
F\left( T_{1},T_{2}\right) \overset{df}{=}\gamma f_{1}\left( T_{1}\right)
-\gamma f_{3}\left( T_{2}\right) -\left( T_{2}-T_{1}\right) ^{2}+\left(
T_{2}-T_{1}\right) V_{1}=0 \\ 
G\left( T_{1},T_{2}\right) \overset{df}{=}\gamma f_{3}\left( T_{2}\right)
-\gamma f_{1}\left( T_{1}\right) -\left( T_{1}+1-T_{2}\right) ^{2}+\left(
T_{1}+1-T_{2}\right) V_{2}=0%
\end{array}
\label{T1T2}
\end{equation}

We can, in principle, solve the equation $F\left( T_{1},T_{2}\right) =0$ to
determine $T_{2}\left( T_{1}\right) $ and demand that $\frac{d}{dT_{1}}%
G\left( T_{1},T_{2}\left( T_{1}\right) \right) =0$ to obtain condition for
double root:%
\begin{eqnarray}
F\left( T_{1},T_{2}\right) &=&0  \notag \\
G\left( T_{1},T_{2}\right) &=&0  \label{cond1} \\
\tfrac{d}{dT_{1}}G\left( T_{1},T_{2}\right) &=&\tfrac{\partial }{\partial
T_{1}}G\left( T_{1},T_{2}\right) +\tfrac{\partial }{\partial T_{1}}G\left(
T_{1},T_{2}\right) \tfrac{dT_{2}}{dT_{1}}=0  \notag
\end{eqnarray}%
where the derivative $\tfrac{dT_{2}}{dT_{1}}$ is computed from Eqn. (\ref%
{Impl1b}). Eqns. (\ref{cond1}) provide analytical condition for the onset of
the $2$ -- cycle. They are too complicated to be solved analytically but can
be solved numerically for a fixed value of $R$ or $\gamma $. For example,
for $R=0.85$ we compute the critical value of $\gamma $ and the critical $2$
-- cycle: $T_{1}=8.\,\allowbreak 167\,748\,882\,\allowbreak
344\,294\,132\,\allowbreak 7\times 10^{-2}$, $T_{2}=0.715\,957\,449\,%
\allowbreak 501\,190\,297\,\allowbreak 28$, $\gamma _{cr,C}^{\left( 2\right)
}=3.\,\allowbreak 661\,705\,268\,\allowbreak 289\,225\,062\,\allowbreak
0\times 10^{-2}$ in perfect agreement with solution of Eqns. (\ref{cr1}), (%
\ref{cr2}).

\subsection{Low velocity $2$ - cycle, model $\mathcal{M}_{S}$}

We can apply this result to the case of sinusoidal motion. First of all,
there is analogous $2$ - cycle with $T_{1}\in \left( 0,\ \frac{1}{4}\right)
,\ T_{2}\in \left( \frac{1}{2},\ \frac{3}{4}\right) $, which appears at $%
\gamma _{cr,S}^{\left( 2\right) }\cong 0.0346$, see Fig. 2. It can be thus
assumed that this $2$ - cycle is also born as a double solution. The
corresponding equations of the $2$ - cycle\ are of form:%
\begin{equation}
\begin{array}{rcl}
\gamma \sin \left( 2\pi T_{2}\right) & = & \gamma \sin \left( 2\pi
T_{1}\right) -\left( T_{2}-T_{1}\right) ^{2}+\left( T_{2}-T_{1}\right) V_{1}
\\ 
V_{2} & = & -RV_{1}+2R\left( T_{2}-T_{1}\right) +\gamma \left( 1+R\right)
2\pi \cos \left( 2\pi T_{2}\right) \\ 
\gamma \sin \left( 2\pi T_{1}\right) & = & \gamma \sin \left( 2\pi
T_{2}\right) -\left( T_{1}+1-T_{2}\right) ^{2}+\left( T_{1}+1-T_{2}\right)
V_{2} \\ 
V_{1} & = & -RV_{2}+2R\left( T_{1}+1-T_{2}\right) +\gamma \left( 1+R\right)
2\pi \cos \left( 2\pi T_{1}\right)%
\end{array}
\label{2-cycle-sine}
\end{equation}%
Solving second and fourth equations of (\ref{2-cycle-sine}) for $V_{1}$, $%
V_{2}$ we get:%
\begin{equation}
\begin{array}{l}
V_{1}=\frac{2}{1-R}\left( RT_{1}-RT_{2}+\gamma \pi \cos \left( 2\pi
T_{1}\right) -\gamma R\pi \cos \left( 2\pi T_{2}\right) +\frac{R}{1+R}\right)
\\ 
V_{2}=\frac{2}{1-R}\left( -RT_{1}+RT_{2}-\gamma R\pi \cos \left( 2\pi
T_{1}\right) +\gamma \pi \cos \left( 2\pi T_{2}\right) -\frac{R^{2}}{1+R}%
\right)%
\end{array}
\label{V1V2sine}
\end{equation}%
The problem is thus reduced to the system of two equations for $T_{1}$, $%
T_{2}$%
\begin{equation}
\begin{array}{cl}
F\left( T_{1},T_{2}\right) \overset{df}{=} & \gamma \sin \left( 2\pi
T_{1}\right) -\gamma \sin \left( 2\pi T_{2}\right) -\Delta ^{2}+V_{1}\Delta
=0 \\ 
G\left( T_{1},T_{2}\right) \overset{df}{=} & \gamma \sin \left( 2\pi
T_{2}\right) -\gamma \sin \left( 2\pi T_{1}\right) -\tilde{\Delta}^{2}+V_{2}%
\tilde{\Delta}=0 \\ 
& \Delta =T_{2}-T_{1},\ \tilde{\Delta}=T_{1}+1-T_{2}%
\end{array}
\label{T1T2sine}
\end{equation}

We couldn't solve the system of equations (\ref{V1V2sine}), (\ref{T1T2sine})
analytically. Analytical condition for double root of these equations, i.e.
for the beginning of the $2$ - cycle, are again provided by Eqns. (\ref%
{cond1}), (\ref{Impl1b}) with functions $F$, $G$ defined in (\ref{T1T2sine}%
). Solving now these equations numerically for $R=0.85$ we get the critical
value of the parameter $\gamma $ and values of dynamical variables of the
critical $2$ - cycle: $\gamma _{cr,S}^{\left( 2\right) }=3.\,\allowbreak
458\,072\,636\,\allowbreak 337\,462\,017\,\allowbreak 6\times 10^{-2}$, $%
T_{1}=7.\,\allowbreak 171\,236\,860\,\allowbreak 717\,641\,004\,\allowbreak
3\times 10^{-2}$, $T_{2}=0.706\,981\,761\,\allowbreak
358\,463\,856\,\allowbreak 03$.

Numerical computations show that at $\gamma =\gamma _{cr,S}^{\left( 2\right)
}$ there is indeed a double solution of (\ref{2-cycle-sine}), two real
solutions for $\gamma >\gamma _{cr,S}^{\left( 2\right) }$ (one stable,
another unstable) and complex solutions for $\gamma <\tilde{\gamma}%
_{cr}^{\left( 2\right) }$. These considerations describe and explain the
birth of the corresponding $2$ - cycles.

\subsection{Low velocity $3$ - cycle in the model $\mathcal{M}_{C}$}

We have found numerically that a $3$ - cycle satisfying conditions $T_{1}\in
\left( 0,\ \frac{1}{4}\right) ,\ T_{2}\in \left( \frac{1}{2},\ \frac{3}{4}%
\right) ,$ $T_{3}\in \left( \frac{3}{4},\ 1\right) $ and $T_{4}=T_{1}+1$\ is
stable. This attractor is seen in the bifurcation diagram near the $2$ -
cycle for $\gamma \gtrsim 0.0452$, $R=0.85$, cf. Fig. 1. The $3$ - cycle
variables fulfill equations:%
\begin{equation}
\begin{array}{rll}
\gamma f_{3}\left( T_{2}\right) & = & \gamma f_{1}\left( T_{1}\right)
-\left( T_{2}-T_{1}\right) ^{2}+\left( T_{2}-T_{1}\right) V_{1} \\ 
V_{2} & = & -RV_{1}+2R\left( T_{2}-T_{1}\right) +\gamma \left( 1+R\right)
g_{3}\left( T_{2}\right) \\ 
\gamma f_{4}\left( T_{3}\right) & = & \gamma f_{3}\left( T_{2}\right)
-\left( T_{3}-T_{2}\right) ^{2}+\left( T_{3}-T_{2}\right) V_{2} \\ 
V_{3} & = & -RV_{2}+2R\left( T_{3}-T_{2}\right) +\gamma \left( 1+R\right)
g_{4}\left( T_{3}\right) \\ 
\gamma f_{1}\left( T_{4}-1\right) & = & \gamma f_{4}\left( T_{3}\right)
-\left( T_{4}-T_{3}\right) ^{2}+\left( T_{4}-T_{3}\right) V_{3} \\ 
V_{4} & = & -RV_{3}+2R\left( T_{4}-T_{3}\right) +\gamma \left( 1+R\right)
g_{1}\left( T_{4}-1\right) \\ 
T_{4} & = & T_{1}+1 \\ 
V_{4} & = & V_{1}%
\end{array}
\label{3-cycle-1}
\end{equation}

Equations (\ref{3-cycle-1}) can be simplified. We can solve the second,
fourth and sixth equations for $V_{1}$, $V_{2}$, $V_{3}$ to get%
\begin{equation}
\begin{array}{l}
V_{1}=\Gamma g_{1}\left( T_{1}\right) +\Gamma R^{2}g_{3}\left( T_{2}\right)
-\Gamma Rg_{4}\left( T_{3}\right) +aT_{1}+bT_{2}-cT_{3}+d \\ 
V_{2}=-\Gamma Rg_{1}\left( T_{1}\right) +\Gamma g_{3}\left( T_{2}\right)
+\Gamma R^{2}g_{4}\left( T_{3}\right) -cT_{1}+aT_{2}+bT_{3}-dR \\ 
V_{3}=\Gamma R^{2}g_{1}\left( T_{1}\right) -\Gamma Rg_{3}\left( T_{2}\right)
+\Gamma g_{4}\left( T_{3}\right) +bT_{1}-cT_{2}+aT_{3}+dR^{2} \\ 
\Gamma =\frac{\gamma }{R^{2}-R+1},\ a=\frac{2R\left( 1-R\right) }{R^{2}-R+1}%
,\ b=\frac{2R^{2}}{R^{2}-R+1},\ c=\frac{2R}{R^{2}-R+1},\ d=\frac{2R}{1+R^{3}}%
\end{array}
\label{V1V2V3}
\end{equation}%
The problem is thus reduced to three equations for impact times $T_{1}$, $%
T_{2}$, $T_{3}$ only:

\begin{equation}
\begin{array}{ll}
F\left( T_{1},T_{2},T_{3}\right) \overset{df}{=} & \gamma f_{1}\left(
T_{1}\right) -\gamma f_{3}\left( T_{2}\right) -\Delta _{1}^{2}+\Delta
_{1}V_{1}=0 \\ 
G\left( T_{1},T_{2},T_{3}\right) \overset{df}{=} & \gamma f_{3}\left(
T_{2}\right) -\gamma f_{4}\left( T_{3}\right) -\Delta _{2}^{2}+\Delta
_{2}V_{2}=0 \\ 
H\left( T_{1},T_{2},T_{3}\right) \overset{df}{=} & \gamma f_{4}\left(
T_{3}\right) -\gamma f_{1}\left( T_{1}\right) -\Delta _{3}^{2}+\Delta
_{3}V_{3}=0 \\ 
& \Delta _{1}=T_{2}-T_{1},\ \Delta _{2}=T_{3}-T_{2},\ \Delta
_{3}=T_{1}+1-T_{3}%
\end{array}
\label{T1T2T3}
\end{equation}%
where $V_{1}$, $V_{2}$, $V_{3}$ are known functions of impact times, cf. (%
\ref{V1V2V3}). We were unable to solve Eqns. (\ref{3-cycle-1}) analytically.
However, it is possible to write down condition for the onset of the $3$ -
cycle since it follows from numerical computations that the $3$ -- cycle is
born as a double root of Eqns. (\ref{3-cycle-1}). The condition for the
double root is $\tfrac{d}{dT_{1}}H\left( T_{1},T_{2}\left( T_{1}\right)
,T_{3}\left( T_{1}\right) \right) =0$ and hence the condition for the onset
of the $3$ -- cycle is:%
\begin{eqnarray}
F\left( T_{1},T_{2},T_{3}\right) &=&0  \notag \\
G\left( T_{1},T_{2},T_{3}\right) &=&0  \label{cond2} \\
H\left( T_{1},T_{2},T_{3}\right) &=&0  \notag \\
\tfrac{d}{dT_{1}}H\left( T_{1},T_{2},T_{3}\right) &=&\tfrac{\partial H}{%
\partial T_{1}}+\tfrac{\partial H}{\partial T_{2}}\tfrac{\partial T_{2}}{%
\partial T_{1}}+\tfrac{\partial H}{\partial T_{3}}\tfrac{\partial T_{3}}{%
\partial T_{1}}=0  \notag
\end{eqnarray}%
where the derivatives $\tfrac{\partial T_{2}}{\partial T_{1}}$, $\tfrac{%
\partial T_{3}}{\partial T_{1}}$ are computed from (\ref{Impl2b}). Solving
these equations numerically for $R=0.85$ we get critical value of the
control parameter $\gamma $ and the critical $3$ - cycle: $\gamma
_{cr,C}^{\left( 3\right) }=4.\,\allowbreak 518\,834\,447\,\allowbreak
846\,807\,553\,\allowbreak 9\times 10^{-2}$, $T_{1}=0.103\,931\,597\,153%
\,962\,754\,\allowbreak 97$, $T_{2}=0.635\,200\,266\,\allowbreak
830\,198\,212\,\allowbreak 15$, $T_{3}=0.848\,760\,321\,\allowbreak
631\,572\,414\,\allowbreak 99$.

\subsection{Low velocity $3$ - cycle in the model $\mathcal{M}_{S}$}

We can apply this result to the case of sinusoidal motion. First of all,
there is analogous $3$ - cycle with $T_{1}\in \left( 0,\ \frac{1}{4}\right)
,\ T_{2}\in \left( \frac{1}{2},\ \frac{3}{4}\right) ,$ $T_{3}\in \left( 
\frac{3}{4},\ 1\right) $ and $T_{4}=T_{1}+1$, which appears at $\tilde{\gamma%
}_{cr}^{\left( 3\right) }\cong 0.04499$, see Fig. 2. We can thus expect that
this $3$ - cycle is also born as a double solution. Dynamical variables of
the $3$ - cycle obey equations:

\begin{equation}
\begin{array}{rcl}
\gamma \sin \left( 2\pi T_{2}\right) & = & \gamma \sin \left( 2\pi
T_{1}\right) -\left( T_{2}-T_{1}\right) ^{2}+\left( T_{2}-T_{1}\right) V_{1}
\\ 
V_{2} & = & -RV_{1}+2R\left( T_{2}-T_{1}\right) +\gamma \left( 1+R\right)
2\pi \cos \left( 2\pi T_{2}\right) \\ 
\gamma \sin \left( 2\pi T_{3}\right) & = & \gamma \sin \left( 2\pi
T_{2}\right) -\left( T_{3}-T_{2}\right) ^{2}+\left( T_{3}-T_{2}\right) V_{2}
\\ 
V_{3} & = & -RV_{2}+2R\left( T_{3}-T_{2}\right) +\gamma \left( 1+R\right)
2\pi \cos \left( 2\pi T_{3}\right) \\ 
\gamma \sin \left( 2\pi T_{1}\right) & = & \gamma \sin \left( 2\pi
T_{3}\right) -\left( T_{1}+1-T_{3}\right) ^{2}+\left( T_{1}+1-T_{3}\right)
V_{3} \\ 
V_{1} & = & -RV_{3}+2R\left( T_{1}+1-T_{3}\right) +\gamma \left( 1+R\right)
2\pi \cos \left( 2\pi T_{1}\right)%
\end{array}
\label{3-cycle-sine}
\end{equation}

Solving second, fourth and sixth equations for $V_{1}$, $V_{2}$, $V_{3}$ we
get%
\begin{equation}
\begin{array}{l}
V_{1}=a\left( R^{2}C_{2}-RC_{3}+C_{1}\right) +b\left( R\left(
T_{2}-T_{1}\right) +T_{1}-T_{3}\right) +c \\ 
V_{2}=a\left( R^{2}C_{3}-RC_{1}+C_{2}\right) +b\left( R\left(
T_{3}-T_{2}\right) +T_{2}-T_{1}\right) -cR \\ 
V_{3}=a\left( R^{2}C_{1}-RC_{2}+C_{3}\right) +b\left( R\left(
T_{1}-T_{3}\right) +T_{3}-T_{2}\right) +cR^{2} \\ 
a=\frac{2\gamma \pi \left( 1+R\right) }{1+R^{3}},\ b=\frac{2R\left(
1+R\right) }{1+R^{3}},\ c=\frac{2R}{1+R^{3}},\ C_{i}=\cos 2\pi T_{i}\quad
\left( i=1,2,3\right)%
\end{array}
\label{V1V2V3sine}
\end{equation}%
and we have to solve equations for impact times only:

\begin{equation}
\begin{array}{l}
F\left( T_{1},T_{2},T_{3}\right) \overset{df}{=}\gamma \sin \left( 2\pi
T_{1}\right) -\gamma \sin \left( 2\pi T_{2}\right) -\Delta _{1}^{2}+\Delta
_{1}V_{1}=0 \\ 
G\left( T_{1},T_{2},T_{3}\right) \overset{df}{=}\gamma \sin \left( 2\pi
T_{2}\right) -\gamma \sin \left( 2\pi T_{3}\right) -\Delta _{2}^{2}+\Delta
_{2}V_{2}=0 \\ 
H\left( T_{1},T_{2},T_{3}\right) \overset{df}{=}\gamma \sin \left( 2\pi
T_{3}\right) -\gamma \sin \left( 2\pi T_{1}\right) -\Delta _{3}^{2}+\Delta
_{3}V_{3}=0 \\ 
\Delta _{1}=T_{2}-T_{1},\ \Delta _{2}=T_{3}-T_{2},\ \Delta _{3}=T_{1}+1-T_{3}%
\end{array}
\label{T1T2T3sine}
\end{equation}

Equations (\ref{V1V2V3sine}), (\ref{T1T2T3sine}) are too complicated to be
solved analytically. However, it is possible to write down condition for the
beginning of the $3$ - cycle since it follows from numerical computations
that the $3$ - cycle is born as a double root of Eqns. (\ref{3-cycle-sine}).
More exactly, we have to solve Eqns. (\ref{cond2}) for functions $F$, $G$, $%
H $ defined in (\ref{T1T2T3sine}). We thus get for $R=0.85$\ the critical
value $\gamma _{cr,S}^{\left( 3\right) }=4.\,\allowbreak
498\,669\,496\,\allowbreak 445\,746\,754\,\allowbreak 8\times 10^{-2}$ and
the critical $3$ - cycle, $T_{1}=9.\,\allowbreak 514\,258\,132\,\allowbreak
574\,445\,543\,\allowbreak 3\times
10^{-2},T_{2}=0.633\,092\,075\,\allowbreak 481\,873\,314\,\allowbreak
56,T_{3}=0.848\,082\,849\,\allowbreak 264\,211\,982\,\allowbreak 09$.

\section{Birth of high velocity $3$ - cycles}

High velocity $3$ - cycles are very characteristic of bouncing ball
dynamics. They accompany all fixed points and are seen in the bifurcation
diagrams around $V=1,\ 2,\ 3,\ \ldots $ , see Figs. 1, 2. In the case of
such cycles $T_{1},T_{2},T_{3}\in \left( 0,1\right) $ and $T_{4}-k=T_{1}$.

\subsection{Model $\mathcal{M}_{C}$, $V\protect\cong 1$}

We start with such $3$ - cycle with $V\cong 1$\ which appears in the model $%
\mathcal{M}_{C}$ for $\gamma \gtrsim 0.042$, see Fig. 1 with impact times $%
T_{1}\in \left( \frac{1}{4},\frac{1}{2}\right) ,\ T_{2}-1\in \left( 0,\frac{1%
}{4}\right) ,\ T_{3}-1\in \left( 0,\tfrac{1}{4}\right) ,\ T_{4}-1=T_{1}$.
The corresponding equations are:%
\begin{equation}
\begin{array}{rl}
\gamma f_{1}\left( T_{2}-1\right) = & \gamma f_{2}\left( T_{1}\right)
-\left( T_{2}-T_{1}\right) ^{2}+\left( T_{2}-T_{1}\right) V_{1} \\ 
V_{2}= & -RV_{1}+2R\left( T_{2}-T_{1}\right) +\gamma \left( 1+R\right)
g_{1}\left( T_{2}-1\right) \\ 
\gamma f_{1}\left( T_{3}-1\right) = & \gamma f_{1}\left( T_{2}-1\right)
-\left( T_{3}-T_{2}+1\right) ^{2}+\left( T_{3}-T_{2}+1\right) V_{2} \\ 
V_{3}= & -RV_{2}+2R\left( T_{3}-T_{2}+1\right) +\gamma \left( 1+R\right)
g_{1}\left( T_{3}-1\right) \\ 
\gamma f_{2}\left( T_{1}\right) = & \gamma f_{1}\left( T_{3}-1\right)
-\left( T_{1}+2-T_{3}\right) ^{2}+\left( T_{1}+2-T_{3}\right) V_{3} \\ 
V_{1}= & -RV_{3}+2R\left( T_{1}+1-T_{3}+1\right) +\gamma \left( 1+R\right)
g_{2}\left( T_{1}\right)%
\end{array}
\label{big3cycle}
\end{equation}

Solving equations for $V_{1},\ V_{2},\ V_{3}$ we get%
\begin{equation}
\begin{array}{l}
V_{1}=a\gamma g_{2}\left( T_{1}\right) -aR\gamma g_{1}\left( T_{3}-1\right)
+aR^{2}\gamma g_{1}\left( T_{2}-1\right) -2aRA_{1}+b \\ 
V_{2}=\gamma ag_{1}\left( T_{2}-1\right) -aR\gamma g_{2}\left( T_{1}\right)
+a\gamma R^{2}g_{1}\left( T_{3}-1\right) -2aRA_{2}-b \\ 
V_{3}=a\gamma g_{1}\left( T_{3}-1\right) -aR\gamma g_{1}\left(
T_{2}-1\right) +a\gamma R^{2}g_{2}\left( T_{1}\right) +2aRA_{3}+c \\ 
A_{1}=-RT_{2}+RT_{1}-T_{1}+T_{3},\ A_{2}=-RT_{3}+RT_{2}-T_{2}+T_{1} \\ 
A_{3}=-RT_{3}+RT_{1}-T_{2}+T_{3} \\ 
a=\frac{\left( 1+R\right) }{1+R^{3}},\ b=\frac{2R\left( 2-R\right) }{1+R^{3}}%
,\ c=\frac{2R\left( 1+2R^{2}\right) }{1+R^{3}}%
\end{array}
\label{V1V2V3big}
\end{equation}%
and%
\begin{equation}
\begin{array}{l}
F_{1}\left( T_{1},T_{2},T_{3}\right) \overset{df}{=}\gamma f_{2}\left(
T_{1}\right) -\gamma f_{1}\left( T_{2}-1\right) -\Delta _{1}^{2}+\Delta
_{1}V_{1}=0 \\ 
F_{2}\left( T_{1},T_{2},T_{3}\right) \overset{df}{=}\gamma f_{1}\left(
T_{2}-1\right) -\gamma f_{1}\left( T_{3}-1\right) -\Delta _{2}^{2}+\Delta
_{2}V_{2}=0 \\ 
F_{3}\left( T_{1},T_{2},T_{3}\right) \overset{df}{=}\gamma f_{1}\left(
T_{3}-1\right) -\gamma f_{2}\left( T_{1}\right) -\Delta _{3}^{2}+\Delta
_{3}V_{3}=0 \\ 
\Delta _{1}=T_{2}-T_{1},\ \Delta _{2}=T_{3}-T_{2}+1,\ \Delta
_{3}=T_{1}+2-T_{3}%
\end{array}
\label{T1T2T3big}
\end{equation}

Analytical condition for the onset of this $3$ -- cycle is given by Eqns. (%
\ref{cond2}) with $F$, $G$, $H$ given by (\ref{T1T2T3big}). Solving these
equations numerically for $R=0.85$ we get $\gamma _{cr}^{\left( 3,1\right)
}=4.\,\allowbreak 184\,258\,672\,\allowbreak 013\,445\,046\,\allowbreak
3\times 10^{-2}$ and $T_{1}=0.292\,944\,344\,346\,867\,579\,\allowbreak 94$, 
$T_{2}=1.\,\allowbreak 113\,429\,439\,\allowbreak 708\,681\,876\,6$, $%
T_{3}=1.\,\allowbreak 179\,198\,676\,090\,233\,556\,9$. For $%
R_{cr,C}=0.685\,101\,194$ and \textbf{\ }$\gamma _{cr,C}=0.056\,81\,9\,493$
there is smooth transition to the state $T_{1}\in \left( \frac{1}{4},\frac{1%
}{2}\right) ,\ T_{2}\in \left( \frac{3}{4},1\right) ,\ T_{3}-1\in \left( 0,%
\tfrac{1}{4}\right) ,\ T_{4}-1=T_{1}$. For $R>R_{cr,C}$ this transition
occurs for $\gamma >\gamma _{cr,C}$.

\subsection{Model $\mathcal{M}_{S}$, $V\protect\cong 1$}

In the case of sinusoidal motion described by the model $\mathcal{M}_{C}$ we
can see the $3$ -- cycle with $V\cong 1$ in Fig. 2 for $\gamma \gtrsim 0.045$
with impact times $T_{1}\in \left( 0,\frac{1}{4}\right) $,$\ T_{2}-1\in
\left( \frac{1}{4},\frac{1}{2}\right) $,$\ T_{3}-1\in \left( 0,\tfrac{1}{4}%
\right) $, $T_{4}-1=T_{1}$. Dynamical equations read:

\begin{equation}
\begin{array}{rl}
\gamma \sin \left( 2\pi T_{2}\right) = & \gamma \sin \left( 2\pi
T_{1}\right) -\left( T_{2}-T_{1}\right) ^{2}+\left( T_{2}-T_{1}\right) V_{1}
\\ 
V_{2}= & -RV_{1}+2R\left( T_{2}-T_{1}\right) +\gamma \left( 1+R\right) 2\pi
\cos \left( 2\pi T_{2}\right) \\ 
\gamma \sin \left( 2\pi T_{3}\right) = & \gamma \sin \left( 2\pi
T_{2}\right) -\left( T_{3}-T_{2}+1\right) ^{2}+\left( T_{3}-T_{2}+1\right)
V_{2} \\ 
V_{3}= & -RV_{2}+2R\left( T_{3}-T_{2}+1\right) +\gamma \left( 1+R\right)
2\pi \cos \left( 2\pi T_{3}\right) \\ 
\gamma \sin \left( 2\pi T_{1}\right) = & \gamma \sin \left( 2\pi
T_{3}\right) -\left( T_{1}+2-T_{3}\right) ^{2}+\left( T_{1}+2-T_{3}\right)
V_{3} \\ 
V_{1}= & -RV_{3}+2R\left( T_{1}+2-T_{3}\right) +\gamma \left( 1+R\right)
2\pi \cos \left( 2\pi T_{1}\right)%
\end{array}
\label{big3-cycle}
\end{equation}

Solving equations for $V_{1},\ V_{2},\ V_{3}$ we get%
\begin{equation}
\begin{array}{l}
V_{1}=aR^{2}\cos 2\pi T_{2}-aR\cos 2\pi T_{3}+a\cos 2\pi T_{1}-bA_{1}-c \\ 
V_{2}=aR^{2}\cos 2\pi T_{3}-aR\cos 2\pi T_{1}+a\cos 2\pi T_{2}-bA_{2}+cR \\ 
V_{3}=aR^{2}\cos 2\pi T_{1}-aR\cos 2\pi T_{2}+a\cos 2\pi T_{3}+bA_{3}+d \\ 
A_{1}=-RT_{2}+RT_{1}-T_{1}+T_{3},\ A_{2}=-RT_{3}+RT_{2}-T_{2}+T_{1} \\ 
A_{3}=RT_{1}-RT_{3}+T_{3}-T_{2} \\ 
a=\frac{2\gamma \pi \left( 1+R\right) }{1+R^{3}},\ b=\frac{2R\left(
1+R\right) }{1+R^{3}},\ c=\frac{2R\left( R-2\right) }{1+R^{3}},\ d=\frac{%
2R\left( 1+2R^{2}\right) }{1+R^{3}}%
\end{array}
\label{V1V2V3bigsine}
\end{equation}%
and%
\begin{equation}
\begin{array}{l}
F\left( T_{1},T_{2},T_{3}\right) \overset{df}{=}\gamma \sin \left( 2\pi
T_{1}\right) -\gamma \sin \left( 2\pi T_{2}\right) -\Delta _{1}^{2}+\Delta
_{1}V_{1}=0 \\ 
G\left( T_{1},T_{2},T_{3}\right) \overset{df}{=}\gamma \sin \left( 2\pi
T_{2}\right) -\gamma \sin \left( 2\pi T_{3}\right) -\Delta _{2}^{2}+\Delta
_{2}V_{2}=0 \\ 
H\left( T_{1},T_{2},T_{3}\right) \overset{df}{=}\gamma \sin \left( 2\pi
T_{3}\right) -\gamma \sin \left( 2\pi T_{1}\right) -\Delta _{3}^{2}+\Delta
_{3}V_{3}=0 \\ 
\Delta _{1}=T_{2}-T_{1},\ \Delta _{2}=T_{3}-T_{2}+1,\ \Delta
_{3}=T_{1}-T_{3}+2%
\end{array}
\label{T1T2T3bigsine}
\end{equation}

Conditions for the onset of $3$ -- cycle are given by Eqns. (\ref{cond2})
with functions $F$, $G$, $H$ defined in (\ref{T1T2T3bigsine}). Solving these
equations for $R=0.85$ we obtain: $\tilde{\gamma}_{cr}^{\left( 3,1\right)
}=4.\,\allowbreak 514\,020\,805\,\allowbreak 615\,479\,834\,\allowbreak
1\times 10^{-2}$ and $T_{1}=7.\,\allowbreak 439\,906\,099\,\allowbreak
929\,247\,941\,\allowbreak 1\times 10^{-2}$, $T_{2}=1.\,\allowbreak
154\,226\,438\,\allowbreak 813\,261\,052\,\allowbreak 6$, $%
T_{3}=1.\,\allowbreak 357\,478\,350\,\allowbreak 324\,075\,728\,\allowbreak
9 $. For $R_{cr,S}=0.691\,964\,922\,5$ and \textbf{\ }$\gamma
_{cr,S}=0.055\,974\,756$ there is smooth transition to the state $T_{1}\in
\left( \frac{1}{4},\frac{1}{2}\right) ,\ T_{2}\in \left( \frac{3}{4}%
,1\right) ,\ T_{3}-1\in \left( 0,\tfrac{1}{4}\right) ,\ T_{4}-1=T_{1}$. For $%
R>R_{cr,S}$ this transition occurs for $\gamma >\gamma _{cr,S}$.

\section{$N$ impacts in one period of limiter's motion and chattering in the
model $\mathcal{M}_{C}$}

In the bouncing ball dynamics chattering and chaotic dynamics arise
typically, see \cite{Giusepponi2003,Giusepponi2005} where chattering
mechanism was studied numerically for sinusoidal motion of the table. Due to
simplicity of our model analytical computations are possible.

We shall consider a possible course of events after grazing.

\subsection{First interval: $T_{i}$,$\ T_{i+1}\in \left( 0,\frac{1}{4}%
\right) $}

Let $T_{i}$,$\ T_{i+1}\in \left( 0,\frac{1}{4}\right) $. In this case we get
from Eqns. (\ref{T}), (\ref{C2b1}) $\Delta _{i+1}=0$ and:%
\begin{equation}
\begin{array}{c}
\Delta _{i+1}^{\left( \pm \right) }=\dfrac{\frac{1}{2}\gamma a_{1}\left(
T_{i}\right) +1}{64\gamma \left( 4-\pi \right) }\left( 1\pm \sqrt{1-\dfrac{%
128\gamma \left( 4-\pi \right) W_{i}}{\left( \frac{1}{2}\gamma a_{1}\left(
T_{i}\right) +1\right) ^{2}}}\right) , \\ 
a_{1}\left( T\right) =\frac{d^{2}}{dT^{2}}f_{1}\left( T\right) ,%
\end{array}
\label{delta1}
\end{equation}%
and $\Delta _{i+1}^{\left( -\right) }$ is the solution describing chattering
(obviously, $W_{i}$ must be small enough so that expression under the square
root be non-negative). The denominator in (\ref{delta1}) can be written as $%
\frac{1}{2}\gamma a_{1}\left( T_{i}\right) +1=-96\gamma \left( 4-\pi \right)
\left( T_{i}-T_{cr}^{\left( 1\right) }\right) $ where:%
\begin{equation}
T_{cr}^{\left( 1\right) }=\dfrac{-1+16\gamma \left( -3+\pi \right) }{%
96\gamma \left( -4+\pi \right) },  \label{Tkr1}
\end{equation}%
and we check that $T_{cr}^{\left( 1\right) }\leq \frac{1}{4}$ occurs for $%
\gamma \geq \gamma _{cr}^{\left( 1\right) }=0.043\,731$. Therefore for $%
\gamma <\gamma _{cr}$ the grazing ball will stay on the table forever. Let
us assume that a ball sticks to the table for some time $T_{g}<T_{cr}^{%
\left( 1\right) }\leq \frac{1}{4}$. At critical point $T_{i}=T_{cr}^{\left(
1\right) }$ and $V_{i}=\gamma g_{1}\left( T_{i}\right) $, equations (\ref{T}%
), (\ref{C2b1}) have the degenerate triple solution $T_{i+1}=T_{cr}^{\left(
1\right) }$. For $\gamma >\gamma _{cr}^{\left( 1\right) }$ and $%
T>T_{cr}^{\left( 1\right) }$ the solution $\Delta _{i+1}^{\left( -\right) }$
is no longer valid since $\frac{1}{2}\gamma a_{1}\left( T_{i}\right) +1<0$
and $\Delta _{i+1}^{\left( -\right) }<0$ what is physically unacceptable.
The solution $\Delta _{i+1}^{\left( +\right) }$ is also unacceptable and
thus the ball has to jump to another time interval, $\left( \frac{1}{4},%
\frac{1}{2}\right) $, $\left( \frac{1}{2},\frac{3}{4}\right) $ or further.
We shall now consider the first possibility. Let us assume that the ball
grazes at $T_{i}=T_{cr}^{\left( 1\right) }$ and thus its velocity is that of
the table, $V_{i}=\gamma g_{1}\left( T_{i}\right) $. We thus have to solve
equation for the jump:%
\begin{equation}
\begin{array}{l}
\gamma f_{2}\left( T_{i+1}\right) =\gamma f_{1}\left( T_{i}\right) -\left(
T_{i+1}-T_{i}\right) ^{2}+\left( T_{i+1}-T_{i}\right) V_{i}, \\ 
T_{i}=T_{cr}^{\left( 1\right) }\in \left( 0,\frac{1}{4}\right) ,\quad
T_{i+1}\in \left( \frac{1}{4},\frac{1}{2}\right) ,\quad V_{i}=\gamma
g_{1}\left( T_{i}\right) .%
\end{array}
\label{jump12}
\end{equation}

The solution of (\ref{jump12}) is%
\begin{equation}
T_{cr}^{\left( 1\longrightarrow 2\right) }=\dfrac{\left( \frac{\pi }{12}-%
\frac{1}{2}\right) \left( 4+2^{\frac{2}{3}}+\sqrt[3]{2}\right) +\frac{1}{2}}{%
-4+\pi }\gamma +\frac{1+2^{\frac{2}{3}}+\sqrt[3]{2}}{96\left( -4+\pi \right)
\gamma }.  \label{Tkr1-2}
\end{equation}%
It follows that the interval $\left( T_{cr}^{\left( 1\right) },\
T_{cr}^{\left( 1\longrightarrow 2\right) }\right) $ is the forbidden zone.
The solution (\ref{Tkr2}) is valid for $\gamma \leq \gamma _{cr}^{\left(
1\longrightarrow 2\right) }=0.057\,102$ since for $\gamma >\gamma
_{cr}^{\left( 2\right) }$ we have $T_{cr}^{\left( 1\longrightarrow 2\right)
}>\frac{1}{2}$ contradicting assumptions. For $\gamma >\gamma _{cr}^{\left(
1\longrightarrow 2\right) }$ we thus have to consider the following equation
for the jump to time interval $\left( \frac{1}{2},\frac{3}{4}\right) $:%
\begin{equation}
\begin{array}{l}
\gamma f_{3}\left( T_{i+1}\right) =\gamma f_{1}\left( T_{i}\right) -\left(
T_{i+1}-T_{i}\right) ^{2}+\left( T_{i+1}-T_{i}\right) V_{i}, \\ 
T_{i}=T_{cr}^{\left( 1\right) }\in \left( 0,\frac{1}{4}\right) ,\quad
T_{i+1}\in \left( \frac{1}{2},\frac{3}{4}\right) ,\quad V_{i}=\gamma
g_{1}\left( T_{i}\right) .%
\end{array}
\label{jump13}
\end{equation}

Solution of Eqn. (\ref{jump13}), $X=T_{cr}^{\left( 1\longrightarrow 3\right)
}$, fulfills the following cubic equation:%
\begin{equation}
\begin{array}{l}
a_{0}X^{3}+\left( b_{0}+b_{1}\gamma \right) X^{2}+\left( c_{0}+c_{1}\gamma
+c_{2}\gamma ^{2}\right) X+d_{0}+d_{1}\gamma +d_{2}\gamma ^{2}+d_{3}\gamma
^{3}=0 \\ 
a_{0}=32,\ b_{0}=1,\ b_{1}=-240+64\pi ,\ c_{0}=\frac{1}{96},\ c_{1}=1-\frac{1%
}{3}\pi \\ 
\ c_{2}=304\pi -552-\frac{124}{3}\pi ^{2},\ d_{0}=\frac{1}{27\,648},\ d_{1}=-%
\frac{1}{576}\pi +\frac{1}{192},\  \\ 
d_{2}=\frac{1}{36}\pi ^{2}-\frac{1}{6}\pi +\frac{1}{4},\ d_{3}=\frac{239}{27}%
\pi ^{3}-\frac{296}{3}\pi ^{2}+364\pi -444%
\end{array}
\label{Tkr1-3}
\end{equation}%
with $T_{cr}^{\left( 1\longrightarrow 3\right) }=\frac{X}{\gamma \left(
-4+\pi \right) }$. Eqn. (\ref{Tkr3}) has acceptable solutions, i.e. such
that $T_{cr}^{\left( 1\longrightarrow 3\right) }\in \left( \frac{1}{2},\frac{%
3}{4}\right) $,\ for $\gamma \leq \gamma _{cr}^{\left( 1\longrightarrow
3\right) }=0.087\,308\,825$.

\subsection{Second interval: $T_{i}$,$\ T_{i+1}\in \left( \frac{1}{4},\ 
\frac{1}{2}\right) $}

We have to consider now chattering in the interval $\left( \frac{1}{4},\ 
\frac{1}{2}\right) $. Let us thus consider that $T_{i},T_{i+1}\in \left( 
\frac{1}{4},\ \frac{1}{2}\right) $. It follows from equations (\ref{T}), (%
\ref{C2b2}) that $\Delta _{i+1}=0$ and%
\begin{equation}
\begin{array}{c}
\Delta _{i+1}^{\left( \pm \right) }=\dfrac{\frac{1}{2}\gamma a_{2}\left(
T_{i}\right) +1}{64\gamma \left( 4-\pi \right) }\left( -1\pm \dfrac{\frac{1}{%
2}\gamma a_{2}\left( T_{i}\right) +1}{\left\vert \frac{1}{2}\gamma
a_{2}\left( T_{i}\right) +1\right\vert }\sqrt{1+\dfrac{128\gamma \left(
4-\pi \right) W_{i}}{\left( \frac{1}{2}\gamma a_{2}\left( T_{i}\right)
+1\right) ^{2}}}\right) , \\ 
a_{2}\left( T\right) =\frac{d^{2}}{dT^{2}}f_{2}\left( T\right) .%
\end{array}
\label{delta2}
\end{equation}

The solution describing chattering is $\Delta _{i+1}^{\left( +\right) }$ for 
$\frac{1}{2}\gamma a_{2}\left( T_{i}\right) +1>0$. The denominator can be
written in form $\frac{1}{2}\gamma a_{2}\left( T_{i}\right) +1=96\gamma
\left( 4-\pi \right) \left( T-T_{cr}^{\left( 2\right) }\right) $ where:%
\begin{equation}
T_{cr}^{\left( 2\right) }=\frac{1}{96}\frac{1-144\gamma +32\gamma \pi }{%
\gamma \left( \pi -4\right) }.  \label{Tkr2}
\end{equation}%
It follows that chattering is thus possible for $T_{i}\in \left(
T_{cr}^{\left( 2\right) },\ \frac{1}{2}\right) $.

\subsection{Third interval: $T_{i},T_{i+1}\in \left( \frac{1}{2},\ \frac{3}{4%
}\right) $}

Let us suppose now that $T_{i},T_{i+1}\in \left( \frac{1}{2},\ \frac{3}{4}%
\right) $. It follows from equations (\ref{T}), (\ref{C2b3}) that $\Delta
_{i+1}=0$ and%
\begin{equation}
\begin{array}{c}
\Delta _{i+1}^{\left( \pm \right) }=\dfrac{\frac{1}{2}\gamma a_{3}\left(
T_{i}\right) +1}{64\gamma \left( 4-\pi \right) }\left( -1\pm \dfrac{\frac{1}{%
2}\gamma a_{3}\left( T_{i}\right) +1}{\left\vert \frac{1}{2}\gamma
a_{3}\left( T_{i}\right) +1\right\vert }\sqrt{1+\dfrac{128\gamma \left(
4-\pi \right) W_{i}}{\left( \frac{1}{2}\gamma a_{3}\left( T_{i}\right)
+1\right) ^{2}}}\right) , \\ 
a_{3}\left( T\right) =\frac{d^{2}}{dT^{2}}f_{3}\left( T\right) .%
\end{array}
\label{delta3}
\end{equation}

The solution describing chattering is $\Delta _{i+1}^{\left( +\right) }$ for 
$\frac{1}{2}\gamma a_{3}\left( T_{i}\right) +1>0$. The denominator can be
written in form $\frac{1}{2}\gamma a_{3}\left( T_{i}\right) +1=96\gamma
\left( 4-\pi \right) \left( T-T_{cr}^{\left( 3\right) }\right) $ with:%
\begin{equation}
T_{cr}^{\left( 3\right) }=\frac{1}{96}\frac{-240\gamma +64\gamma \pi +1}{%
\gamma \left( \pi -4\right) }<0.5\qquad \left( \gamma >0\right) ,
\label{Tkr3}
\end{equation}%
and it follows that $T_{cr}^{\left( 3\right) }$ cannot belong to $\left( 
\frac{1}{2},\ \frac{3}{4}\right) $ interval for positive $\gamma $.
Chattering is thus possible in the whole interval $\left( \frac{1}{2},\ 
\frac{3}{4}\right) $ since $\frac{1}{2}\gamma a_{3}\left( T_{i}\right) +1>0$.

\subsection{Fourth interval: $T_{i}$,$\ T_{i+1}\in \left( \frac{3}{4}%
,1\right) $}

Let us assume finally that two subsequent impacts occur in the last
quarter-period and $T_{i}$,$\ T_{i+1}\in \left( \frac{3}{4},1\right) $. In
this case the solution $\Delta _{i+1}=0$ of equation (\ref{T}) is always
present and this equation can be easily solved. We thus get from Eqns. (\ref%
{T}), (\ref{C2b4}) $\Delta _{i+1}=0$ and:%
\begin{eqnarray}
\Delta _{i+1}^{\left( \pm \right) } &=&\dfrac{\frac{1}{2}\gamma a_{4}\left(
T_{i}\right) +1}{64\gamma \left( 4-\pi \right) }\left( 1\pm \sqrt{1-\dfrac{%
128\gamma \left( 4-\pi \right) W_{i}}{\left( \frac{1}{2}\gamma a_{4}\left(
T_{i}\right) +1\right) ^{2}}}\right) ,  \label{delta4} \\
a_{4}\left( T_{i}\right) &=&\frac{d^{2}}{dT^{2}}f_{4}\left( T\right) ,
\label{a4}
\end{eqnarray}%
where $\gamma a_{4}\left( T_{i}\right) $ is the acceleration of the table, $%
a_{4}\left( T\right) =\frac{d^{2}}{dT^{2}}f_{4}\left( T\right) $ and $W_{i}$
is a relative velocity, $W_{i}=V_{i}-\gamma \frac{d}{dT_{i}}f_{4}\left(
T_{i}\right) $. In the case of chattering the appropriate solution is $%
\Delta _{i+1}^{\left( -\right) }$ since for $W_{i}\longrightarrow 0$ we have 
$\Delta _{i+1}^{\left( -\right) }\longrightarrow 0$. The denominator can be
written as $\frac{1}{2}\gamma a_{4}\left( T_{i}\right) +1=-96\gamma \left(
4-\pi \right) \left( T-T_{cr}^{\left( 4\right) }\right) $ where%
\begin{equation}
T_{cr}^{\left( 4\right) }=\frac{1}{96}\frac{16\gamma \left( 21-5\pi \right)
+1}{\gamma \left( 4-\pi \right) }>1\qquad \left( \gamma >0\right) .
\label{Tkr4}
\end{equation}

We do not have to worry that the denominator in (\ref{delta4}) may vanish
since the condition $\frac{1}{2}\gamma a_{4}\left( T_{i}\right) +1=0$ cannot
be fulfilled for $T_{i}\in \left( \frac{3}{4},1\right) $ and $\gamma >0$.
Therefore chattering is possible in the whole interval $\left( \frac{1}{2},\ 
\frac{3}{4}\right) $ since $\frac{1}{2}\gamma a_{4}\left( T_{i}\right) +1>0$.

\subsection{Grazing: a homoclinic orbit}

Let us assume that the ball grazes at $T_{i}=T_{cr}^{\left( 1\right) }$ with
velocity $V_{i}=\gamma g_{1}\left( T_{i}\right) $ (i.e. it has velocity of
the table) and that the value of $\gamma $ is such that the ball jumps. Let
us next assume that the ball grazes at some time $T_{\ast }$ in the interval 
$\left[ 0,\ T_{cr}^{\left( 1\right) }\right] $. For growing $\gamma $ it
will happen eventually at $\gamma =\gamma _{\ast }$ that $T_{\ast
}=T_{cr}^{\left( 1\right) }$. Then for larger values of $\gamma $ the ball
after grazing and jumping returns with chattering into the $\left( 0,\
T_{cr}^{\left( 1\right) }\right) $ interval but it will not graze, i.e. $%
V_{\ast }$ will be larger than $\gamma g_{1}\left( T_{\ast }\right) $. We
have computed numerically the critical value as $\gamma _{\ast
}=0.058\,348\,6$. Therefore for $\gamma >\gamma _{\ast }$ long transients
can be expected after grazing.

\begin{figure}[th!]
\center 
\includegraphics[width=10cm, height=8cm]{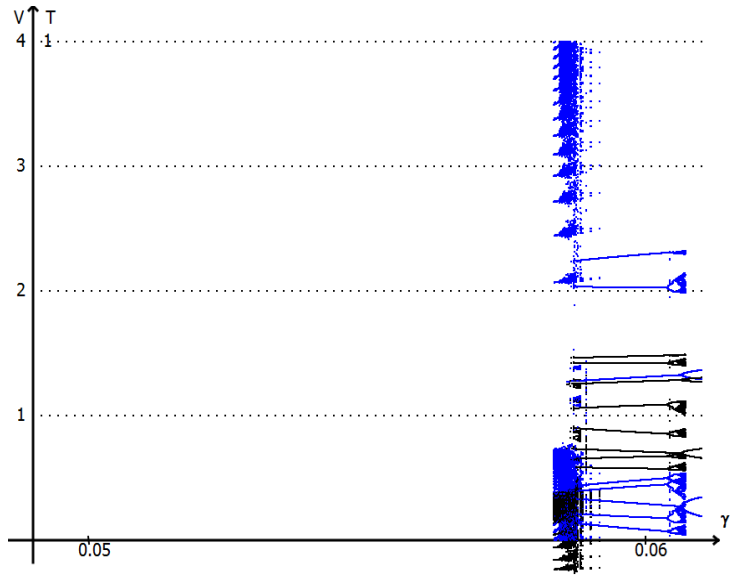}
\caption{Bifurcation diagram for the model $\mathcal{M}_{C}$, $R=0.85$.
Initial conditions are on the grazing manifold.}
\label{F5}
\end{figure}

In Fig. \ref{F5} below bifurcation diagram is shown with initial conditions
on the grazing manifold. For $\gamma <\gamma _{\ast }$ the grazing manifold
is globally attractive. Indeed, in this parameter range the bifurcation diagram 
is empty (we show attractors different than the grazing manifold only).  
On the other hand, for $\gamma >\gamma _{\ast }$ the
ball jumps at $T=T_{cr}^{\left( 1\right) }$ and then either grazes
eventually or settles on some attractor after a long transient. Just after 
the threshold there is a very irregular,  probably chaotic attractor.

\newpage

\section{Summary}

We have studied dynamics of a bouncing ball impacting with a periodically
moving limiter within two frameworks of the table motion: $\mathcal{M}_{C}$
and $\mathcal{M}_{S}$ defined in Section 2. Stability conditions of
fixed points have been determined and results for the models $\mathcal{M}
_{C} $ and $\mathcal{M}_{S}$ have been compared. Then we have found that
low-velocity $k$-cycles as well as high-velocity $3$-cycles are generically
born in tangent bifurcations. Moreover, we have been able using implicit
functions theorems, to write down conditions for the onset of these cycles
and solve them numerically. Analytical conditions for the onset of such cycles 
are new. 

Finally, the case of $N$ impacts in one interval of the limiter's motion has
been studied within the $\mathcal{M}_{C}$ model. Equations for $N$
impacts in one period of limiter's motion were found and simplified
significantly, making analysis of chattering and grazing possible. We have 
found, combining analytical and numerical approach, the grazing homoclinic orbit 
which appears at $\gamma = \gamma _{\ast }$ and gives rise to a very irregular, probably 
chaotic attractor, cf.  Fig. \ref{F5}. We expect that analogous attractor exists 
in the model  $\mathcal{M}_{S}$.

\newpage

\appendix{}

\section{Equation for the $2$ - cycle}

In the Appendix coefficients of the polynomial (\ref{Delta}) are listed.

\bigskip

\noindent $d_{9}=2^{16}3^{3}\gamma ^{3}\left( -4+\pi \right) ^{4}\left(
1+R\right) ^{6}$

\noindent $d_{8}=-2^{12}3^{3}\gamma ^{2}\left( 1+R\right) ^{6}\left( -4+\pi
\right) ^{3}\left( \left( 72\pi -288\right) \gamma +1\right) $

\noindent $d_{7}=2^{6}3^{3}\gamma \left( 1+R\right) ^{4}\left( -4+\pi
\right) ^{2}\left( a_{2}R^{2}+a_{1}R+a_{0}\right) $%
\begin{eqnarray*}
a_{2} &=&\left( -66\,048\pi +8256\pi ^{2}+132\,096\right) \gamma ^{2}+\left(
256\pi -1024\right) \gamma +1 \\
a_{1} &=&\left( -138\,240\pi +17\,280\pi ^{2}+276\,480\right) \gamma
^{2}+2+\left( 512\pi -2048\right) \gamma \\
a_{0} &=&\left( -66\,048\pi +8256\pi ^{2}+132\,096\right) \gamma ^{2}+\left(
256\pi -1024\right) \gamma +1
\end{eqnarray*}%
$d_{6}=-8\left( 1+R\right) ^{4}\left( -4+\pi \right) \left(
b_{2}R^{2}+b_{1}R+b_{0}\right) $%
\begin{eqnarray*}
b_{2} &=&\left( -19\,160\,064\pi ^{2}+76\,640\,256\pi +1596\,672\pi
^{3}-102\,187\,008\right) \gamma + \\
&&\left( -3024+756\pi \right) \gamma +\left( 1361\,664-679\,680\pi
+84\,864\pi ^{2}\right) \gamma ^{2}+1 \\
b_{1} &=&\left( -241\,532\,928+3773\,952\pi ^{3}+181\,149\,696\pi
-45\,287\,424\pi ^{2}\right) \gamma ^{3}+ \\
&&\left( 183\,552\pi ^{2}-1469\,952\pi +2944\,512\right) \gamma ^{2}+\left(
1512\pi -6048\right) \gamma +2 \\
b_{0} &=&\left( -19\,160\,064\pi ^{2}+76\,640\,256\pi +1596\,672\pi
^{3}-102\,187\,008\right) \gamma ^{3}+ \\
&&\left( 1361\,664-679\,680\pi +84\,864\pi ^{2}\right) \gamma ^{2}+\left(
-3024+756\pi \right) \gamma +1
\end{eqnarray*}%
$d_{5}=8\left( 1+R\right) ^{4}\left( c_{2}R^{2}+c_{1}R+c_{0}\right) $%
\begin{eqnarray*}
c_{2} &=&\left( 180\,983\,808-180\,983\,808\pi -11\,319\,552\pi
^{3}+707\,968\pi ^{4}+67\,885\,056\pi ^{2}\right) \gamma ^{3}+ \\
&&\left( 2951\,424\pi -3953\,664+61\,056\pi ^{3}-734\,976\pi ^{2}\right)
\gamma ^{2}+ \\
&&\left( 17\,928+1118\pi ^{2}-8952\pi \right) \gamma +3\pi -12 \\
c_{1} &=&\left( 2202\,368\pi ^{4}+565\,014\,528-564\,682\,752\pi
+211\,636\,224\pi ^{2}-35\,254\,272\pi ^{3}\right) \gamma ^{3}+ \\
&&\left( 163\,584\pi ^{3}-10\,561\,536-1967\,616\pi ^{2}+7893\,504\pi
\right) \gamma ^{2}+ \\
&&\left( -16\,176\pi +32\,400+2020\pi ^{2}\right) \gamma -24+6\pi \\
c_{0} &=&\left( 706\,432\pi ^{4}+68\,106\,240\pi ^{2}-181\,979\,136\pi
-11\,327\,232\pi ^{3}+182\,310\,912\right) \gamma ^{3}+ \\
&&\left( 2951\,424\pi -3953\,664+61\,056\pi ^{3}-734\,976\pi ^{2}\right)
\gamma ^{2}+ \\
&&\left( 17\,928+1118\pi ^{2}-8952\pi \right) \gamma +3\pi -12
\end{eqnarray*}%
$d_{4}=-4\left( 1+R\right) ^{2}\left(
e_{4}R^{4}+e_{3}R^{3}+e_{2}R^{2}+e_{1}R+e_{0}\right) $%
\begin{eqnarray*}
e_{4} &=&\left( 50\,264\,064-3181\,824\pi ^{3}+201\,344\pi
^{4}-50\,264\,064\pi +18\,929\,664\pi ^{2}\right) \gamma ^{3}+ \\
&&\left( 1854\,720\pi -2515\,968-456\,960\pi ^{2}+37\,632\pi ^{3}\right)
\gamma ^{2}+ \\
&&\left( 29\,160-14\,520\pi +1810\pi ^{2}\right) \gamma -36+9\pi \\
e_{3} &=&\left( -29\,354\,496\pi ^{3}+473\,112\,576+176\,357\,376\pi
^{2}+1834\,496\pi ^{4}-471\,453\,696\pi \right) \gamma ^{3}+ \\
&&\left( -2864\,640\pi ^{2}+11\,566\,080\pi -15\,593\,472+236\,928\pi
^{3}\right) \gamma ^{2}+ \\
&&\left( 6160\pi ^{2}+99\,360-49\,440\pi \right) \gamma +30\pi -120 \\
e_{2} &=&\left( -68\,896\pi ^{4}+20\,570\,112\pi +1156\,800\pi
^{3}-21\,689\,856-7310\,880\pi ^{2}\right) \gamma ^{3}+ \\
&&\left( -27\,039\,744-4981\,248\pi ^{2}+412\,416\pi ^{3}+20\,086\,272\pi
\right) \gamma ^{2}+ \\
&&\left( 140\,400-69\,840\pi +8700\pi ^{2}\right) \gamma +42\pi -168 \\
e_{1} &=&\left( 486\,383\,616+1819\,136\pi ^{4}+178\,569\,216\pi
^{2}-29\,431\,296\pi ^{3}-481\,406\,976\pi \right) \gamma ^{3}+ \\
&&\left( -2864\,640\pi ^{2}+11\,566\,080\pi -15\,593\,472+236\,928\pi
^{3}\right) \gamma ^{2}+ \\
&&\left( 6160\pi ^{2}+99\,360-49\,440\pi \right) \gamma +30\pi -120 \\
e_{0} &=&\left( -3220\,224\pi ^{3}+193\,664\pi ^{4}-55\,240\,704\pi
+20\,035\,584\pi ^{2}+56\,899\,584\right) \gamma ^{3}+ \\
&&\left( 1854\,720\pi -2515\,968-456\,960\pi ^{2}+37\,632\pi ^{3}\right)
\gamma ^{2}+ \\
&&\left( 29\,160-14\,520\pi +1810\pi ^{2}\right) \gamma -36+9\pi
\end{eqnarray*}%
$d_{3}=-16\left( 1+R\right) ^{2}\left(
f_{4}R^{4}+f_{3}R^{3}+f_{2}R^{2}+f_{1}R+f_{0}\right) $%
\begin{eqnarray*}
f_{4} &=&\left( 3905\,280-233\,088\pi ^{3}+14\,000\pi ^{4}-3891\,456\pi
+1437\,264\pi ^{2}\right) \gamma ^{3}+ \\
&&\left( -16\,128\pi +27\,648+3072\pi ^{2}-192\pi ^{3}\right) \gamma ^{2}+ \\
&&\left( -3132+1548\pi -192\pi ^{2}\right) \gamma +8-2\pi \\
f_{3} &=&\left( 340\,032\pi ^{3}-6027\,264-2066\,112\pi ^{2}-21\,504\pi
^{4}+5709\,312\pi \right) \gamma ^{3}+ \\
&&\left( 115\,968\pi ^{2}-479\,232\pi +663\,552-9408\pi ^{3}\right) \gamma
^{2}+ \\
&&\left( -622\pi ^{2}-10\,152+5016\pi \right) \gamma -5\pi +20 \\
f_{2} &=&\left( -68\,896\pi ^{4}+20\,570\,112\pi +1156\,800\pi
^{3}-21\,689\,856-7310\,880\pi ^{2}\right) \gamma ^{3}+ \\
&&\left( 1714\,176+308\,736\pi ^{2}-25\,344\pi ^{3}-1257\,984\pi \right)
\gamma ^{2}+ \\
&&\left( -12\,312+6072\pi -752\pi ^{2}\right) \gamma -6\pi +24 \\
f_{1} &=&\left( -9234\,432-17\,792\pi ^{4}-2600\,640\pi ^{2}+358\,592\pi
^{3}+8114\,688\pi \right) \gamma ^{3}+ \\
&&\left( 115\,968\pi ^{2}-479\,232\pi +663\,552-9408\pi ^{3}\right) \gamma
^{2}+ \\
&&\left( -622\pi ^{2}-10\,152+5016\pi \right) \gamma -5\pi +20 \\
f_{0} &=&+\left( -225\,088\pi ^{3}+15\,600\pi ^{4}-2854\,656\pi
+1206\,864\pi ^{2}+2522\,880\right) \gamma ^{3}+ \\
&&\left( -16\,128\pi +27\,648+3072\pi ^{2}-192\pi ^{3}\right) \gamma ^{2}+ \\
&&\left( -3132+1548\pi -192\pi ^{2}\right) \gamma +8-2\pi
\end{eqnarray*}%
$d_{2}=2\left( 1+R\right) ^{2}\left(
g_{4}R^{4}+g_{3}R^{3}+g_{2}R^{2}+g_{1}R+g_{0}\right) $%
\begin{eqnarray*}
g_{4} &=&\left( 9372\,672-534\,528\pi ^{3}+31\,552\pi ^{4}-9206\,784\pi
+3349\,440\pi ^{2}\right) \gamma ^{3}+ \\
&&\left( -152\,064+2656\pi ^{3}-31\,200\pi ^{2}+120\,384\pi \right) \gamma
^{2}+ \\
&&\left( -3456+1632\pi -196\pi ^{2}\right) \gamma +36-9\pi \\
g_{3} &=&\left( -1544\,448\pi ^{3}+24\,551\,424+9416\,448\pi ^{2}+93\,184\pi
^{4}-25\,049\,088\pi \right) \gamma ^{3}+ \\
&&\left( -114\,432\pi ^{2}+419\,328\pi -497\,664+10\,176\pi ^{3}\right)
\gamma ^{2}+ \\
&&\left( -1192\pi ^{2}-19\,872+9696\pi \right) \gamma -12\pi +48 \\
g_{2} &=&\left( 133\,248\pi ^{4}-25\,214\,976\pi -1969\,920\pi
^{3}+21\,731\,328+10\,696\,320\pi ^{2}\right) \gamma ^{3}+ \\
&&\left( 1838\,592+318\,336\pi ^{2}-25\,728\pi ^{3}-1321\,344\pi \right)
\gamma ^{2}+ \\
&&\left( -12\,096+5760\pi -696\pi ^{2}\right) \gamma -18\pi +72 \\
g_{1} &=&\left( 12\,607\,488+107\,008\pi ^{4}+7425\,792\pi ^{2}-1475\,328\pi
^{3}-16\,091\,136\pi \right) \gamma ^{3}+ \\
&&\left( -114\,432\pi ^{2}+419\,328\pi -497\,664+10\,176\pi ^{3}\right)
\gamma ^{2}+ \\
&&\left( -1192\pi ^{2}-19\,872+9696\pi \right) \gamma -12\pi +48 \\
g_{0} &=&\left( -515\,328\pi ^{3}+35\,392\pi ^{4}-6718\,464\pi +2796\,480\pi
^{2}+6054\,912\right) \gamma ^{3}+ \\
&&\left( 113\,472\pi -138\,240-30\,624\pi ^{2}+2720\pi ^{3}\right) \gamma
^{2}+ \\
&&\left( -3456+1632\pi -196\pi ^{2}\right) \gamma 36-9\pi
\end{eqnarray*}%
$d_{1}=2\left( 1+R\right) ^{2}\left(
h_{4}R^{4}+h_{3}R^{3}+h_{2}R^{2}+h_{1}R+h_{0}\right) $%
\begin{eqnarray*}
h_{4} &=&\left( -57\,024\pi ^{3}+3872\pi ^{4}-746\,496\pi +311\,328\pi
^{2}+663\,552\right) \gamma ^{3}+ \\
&&\left( -13\,824+416\pi ^{3}-4128\pi ^{2}+13\,248\pi \right) \gamma ^{2}+ \\
&&\left( 360\pi -48\pi ^{2}-648\right) \gamma +3\pi -12 \\
h_{3} &=&\left( 148\,224\pi ^{3}-3151\,872-1021\,824\pi ^{2}-7552\pi
^{4}+2985\,984\pi \right) \gamma ^{3}+ \\
&&\left( -216\,576\pi -4800\pi ^{3}+276\,480+56\,064\pi ^{2}\right) \gamma
^{2}+ \\
&&\left( 152\pi ^{2}-1248\pi +2592\right) \gamma \\
h_{2} &=&\left( -25\,920\pi ^{4}+5474\,304\pi +395\,136\pi
^{3}-4976\,640-2223\,936\pi ^{2}\right) \gamma ^{3}+ \\
&&\left( -32\,640\pi ^{2}+132\,480\pi +2688\pi ^{3}-179\,712\right) \gamma
^{2}+ \\
&&\left( 720\pi -96\pi ^{2}-1296\right) \gamma +6\pi -24 \\
h_{1} &=&\left( -497\,664-10\,624\pi ^{4}-579\,456\pi ^{2}+132\,864\pi
^{3}+995\,328\pi \right) \gamma ^{3}+ \\
&&\left( -216\,576\pi -4800\pi ^{3}+276\,480+56\,064\pi ^{2}\right) \gamma
^{2} \\
&&\left( 152\pi ^{2}-1248\pi +2592\right) \gamma \\
h_{0} &=&\left( -57\,024\pi ^{3}+3872\pi ^{4}-746\,496\pi +311\,328\pi
^{2}+663\,552\right) \gamma ^{3}+ \\
&&\left( 20\,160\pi +352\pi ^{3}-27\,648-4704\pi ^{2}\right) \gamma ^{2}+ \\
&&\left( 360\pi -48\pi ^{2}-648\right) \gamma +3\pi -12
\end{eqnarray*}%
$d_{0}=k_{6}R^{6}+k_{5}R^{5}+k_{4}R^{4}+k_{3}R^{3}+k_{2}R^{2}+k_{1}R+k_{0}$%
\begin{eqnarray*}
k_{6} &=&\left( -235\,872\pi ^{2}+608\,256\pi -2528\pi
^{4}-580\,608+40\,128\pi ^{3}\right) \gamma ^{3}+ \\
&&\left( 4272\pi ^{2}+17\,280-400\pi ^{3}-14\,976\pi \right) \gamma ^{2}+ \\
&&\left( 216-120\pi +16\pi ^{2}\right) \gamma -\pi +4 \\
k_{5} &=&\left( -7744\pi ^{4}+114\,048\pi ^{3}-1327\,104+1492\,992\pi
-622\,656\pi ^{2}\right) \gamma ^{3}+ \\
&&\left( -32\pi ^{3}-6912+3456\pi -288\pi ^{2}\right) \gamma ^{2}+ \\
&&\left( -240\pi +32\pi ^{2}+432\right) \gamma \\
k_{4} &=&\left( -7456\pi ^{4}+995\,328\pi -746\,496-478\,368\pi
^{2}+98\,880\pi ^{3}\right) \gamma ^{3}+ \\
&&\left( 18\,432\pi -24\,192+368\pi ^{3}-4560\pi ^{2}\right) \gamma ^{2}+ \\
&&\left( -360\pi +648+48\pi ^{2}\right) \gamma +12-3\pi \\
k_{3} &=&\left( 552\,960\pi -256\,896\pi ^{2}+52\,480\pi
^{3}-442\,368-3968\pi ^{4}\right) \gamma ^{3}+ \\
&&\left( -480\pi +864+64\pi ^{2}\right) \gamma \\
k_{2} &=&\left( -588\,960\pi ^{2}+1492\,992\pi +102\,720\pi ^{3}-6688\pi
^{4}-1410\,048\right) \gamma ^{3}+ \\
&&\left( -4272\pi ^{2}+14\,976\pi +400\pi ^{3}-17\,280\right) \gamma ^{2}+ \\
&&\left( -360\pi +648+48\pi ^{2}\right) \gamma -3\pi +12 \\
k_{1} &=&\left( -7744\pi ^{4}+114\,048\pi ^{3}-1327\,104+1492\,992\pi
-622\,656\pi ^{2}\right) \gamma ^{3}+ \\
&&\left( 288\pi ^{2}-3456\pi +32\pi ^{3}+6912\right) \gamma ^{2}+ \\
&&\left( -240\pi +32\pi ^{2}+432\right) \gamma \\
k_{0} &=&\left( 38\,848\pi ^{3}+442\,368\pi -199\,008\pi
^{2}-359\,424-2784\pi ^{4}\right) \gamma ^{3}+ \\
&&\left( -368\pi ^{3}+4560\pi ^{2}-18\,432\pi +24\,192\right) \gamma ^{2}+ \\
&&\left( 216-120\pi +16\pi ^{2}\right) \gamma +4-\pi
\end{eqnarray*}


\begin{thebibliography}{99}
\bibitem{diBernardo2008} M. di Bernardo, C.J. Budd, A.R. Champneys, P.
Kowalczyk, \textit{Piecewise-Smooth Dynamical Systems. Theory and
Applications}. Series: Applied Mathematical Sciences, vol. 163. Springer,
Berlin (2008).

\bibitem{Luo2006} A.C.J.Luo, \textit{Singularity and Dynamics on
Discontinuous Vector Fields}. Monograph Series on Nonlinear Science and
Complexity, vol. 3. Elsevier, Amsterdam (2006).

\bibitem{Awrejcewicz2003} J. Awrejcewicz, C.-H. Lamarque, \textit{%
Bifurcation and Chaos in Nonsmooth Mechanical Systems}.World Scientific
Series on Nonlinear Science: Series A, vol. 45. World Scientific Publishing,
Singapore (2003).

\bibitem{Filippov1988} A.F. Filippov, \textit{Differential Equations with
Discontinuous Right-Hand Sides}. Kluwer Academic, Dordrecht (1988).

\bibitem{Stronge2000} W.J. Stronge, \textit{Impact mechanics}. Cambridge
University Press, Cambridge (2000).

\bibitem{Mehta1994} A. Mehta (ed.), \textit{Granular Matter: An
Interdisciplinary Approach}. Springer, Berlin (1994).

\bibitem{Knudsen1992} C. Knudsen, R. Feldberg, H. True, Bifurcations and
chaos in a model of a rolling wheel-set. Philos. Trans. R. Soc. Lond. A 
\textbf{338}, 455--469 (1992).

\bibitem{Wiercigroch2008} M. Wiercigroch, A.M. Krivtsov, J. Wojewoda,
Vibrational energy transfer via modulated impacts for percussive drilling,
Journal of Theoretical and Applied Mechanics \textbf{46}, 715--726 (2008).

\bibitem{Awrejcewicz2007} J. Awrejcewicz, G. Kudra, G. Wasilewski,
Experimental and numerical investigation of chaotic regions in the triple
physical pendulum, Nonlinear Dynamics \textbf{50}, 755--766 (2007).

\bibitem{Holmes1982} Holmes, P.J.: The dynamics of repeated impacts with a
sinusoidally vibrating table. J. Sound and Vibration \textbf{84}, 173-189
(1982).

\bibitem{Luo1996} A.C.J. Luo, R.P.S. Han, The dynamics of a bouncing ball
with a sinusoidally vibrating table revisited, Nonlinear Dynamics \textbf{10}%
, 1--18 (1996).

\bibitem{Luo2009a} A.C.J. Luo, Y. Guo, Motion Switching and Chaos of a
Particle in a Generalized Fermi-Acceleration Oscillator, Mathematical
Problems in Engineering, vol. \textbf{2009}, Article ID 298906, 40 pages,
2009.

\bibitem{Nordmark2001} A.B. Nordmark, Existence of periodic orbits in
grazing bifurcations of impacting mechanical oscillator, Nonlinearity 
\textbf{14}, 1517--1542 (2001).

\bibitem{Lenci2000} S. Lenci, G. Rega, Periodic solutions and bifurcations
in an impact inverted pendulum under impulsive excitation, Chaos, Solitions
and Fractals \textbf{11}, 2453--2472 (2000).

\bibitem{AOBR2010} A. Okni\'{n}ski, B. Radziszewski, Simple models of
bouncing ball dynamics and their comparison, arXiv:1002.2448 [nlin.CD]
(2010).

\bibitem{AOBR2009} A. Okni\'nski, B. Radziszewski, Dynamics of impacts with
a table moving with piecewise constant velocity, Nonlinear Dynamics \textbf{%
58}, 515--523 (2009).

\bibitem{AOBR2011} A. Okni\'{n}ski, B. Radziszewski, Chaotic dynamics in a
simple bouncing ball model, Acta Mech. Sinica \textbf{27}, 130--134 (2011),
arXiv:1002.2448 [nlin.CD] (2010).

\bibitem{AOBR2012a} A. Okni\'{n}ski, B. Radziszewski, Simple model of
bouncing ball dynamics: displacement of the table assumed as quadratic
function of time, Nonlinear Dynamics \textbf{67}, 1115---1122 (2012).

\bibitem{AOBR2012b} A. Okni\'{n}ski, B. Radziszewski, Simple model of
bouncing ball dynamics. Displacement of the limiter assumed as a cubic
function of time., Differential Equations and Dynamical Systems \textbf{21},
165--171 (2013).

\bibitem{Pieranski1983} P. Piera\'{n}ski, Jumping particle model. Period
doubling cascade in an experimental system, J. Phys. (Paris) \textbf{44},
573--578 (1983).

\bibitem{Tufillaro1986} N.B. Tufillaro, T.M. Mello, Y.M. Choi, N.B. Albano,
Period doubling boundaries of a bouncing ball, J.Phys (Paris) \textbf{47},
1477--1482 (1986).

\bibitem{Celaschi1987} S. Celaschi, R.L. Zimmerman, Evolution of a
two-parameter chaotic dynamics from universal attractors, Phys. Lett. 
\textbf{120A}, 447--451 (1987).

\bibitem{Pieranski1994} P. Piera\'{n}ski, R. Barberi, \textit{Bouncing Ball
Workbench}, OWN\ Pozna\'{n} (1994).

\bibitem{Eichwald2010} B. Eichwald, M. Argentina, X. Noblin, F. Celestini,
Dynamics of a ball bouncing on a vibrated elastic membrane, Phys. Rev. 
\textbf{E82}, 016203 (2010).

\bibitem{AOBR2007} A. Okni\'{n}ski, B. Radziszewski, Grazing dynamics and
dependence on initial conditions in certain systems with impacts,
arXiv:0706.0257v2 [nlin.CD] (2007).

\bibitem{Luo2009b} A.C.J.Luo, \textit{Discontinuous Dynamical System on
Time-varying Domains}. Series: Nonlinear Physical Science. Higher Education
Press, Beijing and Springer, Dordrecht, Heidelberg, London, New York (2009).

\bibitem{Jury1974} E.I.Jury, \textit{Inners and Stability of Dynamic Systems}%
. Wiley, New York (1974) [2nd edn., Krieger, Malabar, 1982].

\bibitem{Krantz2003} S.G. Krantz, H.R. Parks, \textit{The implicit function
theorem: history, theory, and applications. }Birkh\"{a}user, Boston (2002).

\bibitem{Peitgen1992} H.O. Peitgen, H. J\"{u}rgens, and D. Saupe, \textit{%
Fractals for the Classroom. Part Two: Complex Systems and Mandelbrot Set.},
Springer, New York (1992).

\bibitem{Giusepponi2003} S. Giusepponi, F. Marchesoni, The chattering
dynamics of an ideal bouncing ball, Europhysics Letters \textbf{64}, 36
(2003).

\bibitem{Giusepponi2005} S. Giusepponi, F. Marchesoni, M. Borromeo,
Randomness in the bouncing ball dynamics, Physica A \textbf{351}, 142--158
(2005).
\end{thebibliography}
\end{document}